\documentclass[manuscript]{aastex631}
\usepackage{txfonts}
\usepackage{latexsym,bm}
\usepackage{url}
\usepackage{color}
\usepackage{colortbl}
\usepackage{multirow}
\newcommand{\dee}{\mathrm{d}}
\definecolor{mygray}{gray}{.9}

\newcommand{\txt}[1]{\mathrm{#1}}

\usepackage[titletoc,title]{appendix}
\newcommand{\qinemail}
{qingang@hit.edu.cn}
\newcommand{\hit}{School of Science, Harbin Institute of
	Technology, Shenzhen, 518055, China}
\newcommand{\hitqin}{\hit; \qinemail}
\newcommand{\szlab}{Shenzhen Key Laboratory of Numerical Prediction for Space Storm,
	Harbin Institute of Technology, 
	Shenzhen, 518055, China}
\shorttitle{DIMENSIONLESS QUANTITIES DETERMING INFLUENCE OF SOLAR WIND TO DIFFUSION}
\shortauthors{WANG AND QIN}

\begin{document}
\arraycolsep 0pt

\title{The effect of solar wind on 
the charged particles' diffusion coefficients}

\correspondingauthor{G. Qin}
\email{\qinemail}

\author[0000-0002-9586-093X]{J. F. Wang}
\affiliation{\hitqin}

\author[0000-0002-3437-3716]{G. Qin}
\affiliation{\hitqin}
\affiliation{\szlab}

\begin{abstract}
The transport of energetic 
charged particles
through magnetized plasmas 
is ubiquitous in interplanetary space
and astrophysics, and 
the important 
physical quantities are
the along-field and cross-field 
spatial diffusion coefficients
of energetic charged particles.
In this paper, 
the influence of solar wind
on particle transport 
is investigated.
Using the 
focusing equation, 
we obtain along- and 
cross-field diffusion 
coefficient
accounting for the solar wind effect. 
For different conditions, 
the relative importance 
of solar wind effect to diffusion
are investigated. 
It is shown that  
when energetic charged 
particles are
close to the sun, 
for along-field 
diffusion
the solar wind  effect  
needs to be taken into account.  
These results are important for studying 
energetic charged particle 
transport
processes in the vicinity of 
the sun.  
\end{abstract}

\keywords{Interplanetary turbulence (830); Magnetic fields (994); Solar energetic
particles (1491)}

%%%%%%%%%%%%%%%%%%%%%%%%
\section{INTRODUCTION}

The charged energetic particles 
emitted by the sun, known as 
solar energetic particles (SEPs),  
have crucial impacts on the 
environment of interplanetary and
planetary space
\citep{Schlickeiser2002, Reames2017}. 
For example, SEPs can cause geomagnetic
and ionospheric storms on the Earth,
and even pose a threat to the safe 
operation of the ground power 
systems and cause leaks in underground oil pipelines. In addition,   
these energetic particles can
reduce the reliability of  
spacecraft-borne detectors 
and endanger the health 
of astraurants
and aircrew
\citep{Lanzerotti2017, 
MertensEA2018}.
The turbulent
magnetized plasmas
in the interplanetary space,
e.g., the solar wind, 
have a significant impact on 
the transport of solar energetic
particles. Therefore, 
it is extremely important  
to investigate
the propagation of solar
energetic particles through
the solar wind
\citep{Jokipii1966,
Zhang1999,
Schlickeiser2002, 
MatthaeusEA2003, Qin2007, 
SchlickeiserEA2007, 
SchlickeiserEA2008, Shalchi2010,
qz2014, WangAQin15, 
ZhangandZhao2017, ZhangEA2019,
ZhaoEA2017}.

Due to the interaction with
the background and superposed 
turbulent magnetic fields, 
the motion of charged energetic particles
can be modelled as 
two components of motions,
i.e., the
helical motion around 
the mean
magnetic field lines 
and the superposed stochastic one
\citep{Schlickeiser2002,
Shalchi2009}.  
Therefore,  
various statistical methods
have to be utilized 
in related studies
\citep{Jokipii1966,
Schlickeiser2002, MatthaeusEA2003, 
Shalchi2009,
Shalchi2010, Shalchi2020b}. 
The well-known Master equation
in statistical
physics, which provides the most fundamental 
description of the transport
of charged energetic particles
through magnetized plasmas, 
is too
complicated to be used in
analytical investigations
\citep{Schlickeiser2002, Shalchi2009}.
Therefore, in the previous papers
the relatively simple equations, 
such as 
the Fokker-Planck equation,
have been widely used 
in plasma physics, 
astrophysics and space physics. 
The focusing equation, 
the special version of the
Fokker-Planck equation,
has been extensively employed in the research
of the
energetic particle transport in the 
heliosphere and the 
magnetosphere
\citep{Skilling1971, Schlickeiser2002, 
QinEA2005,
QinEA2006, ZhangEA2009, 
DrogeEA2010, ZuoEA2011, 
WangEA2012, QinEA2013, 
ZuoEA2013, WangEA2014,
ZhangEA2019, QinandQiEA2020}. 

The focusing equation includes 
all the important 
transport effects of SEPs 
in solar wind plasmas, 
e.g., 
the pitch-angle diffusion,
the cross-field diffusion,
the spatial convection,
the adiabatic cooling,
the adiabatic focusing,
and so on, among which,  
the spatial convection,
the adiabatic cooling,
and the adiabatic focusing 
are affected by solar wind velocity effects
\citep{Skilling1971, Schlickeiser2002,
ForbesEA2006, Shalchi2009, ZhangandZhao2017,
WijsenEA2019, ZhangEA2019, 
BianEA2020}.
These effects in the focusing 
equation are not isolated from 
each other, but have mutual 
influence. 
The along- and cross-field spatial 
diffusion are very
important transport processes of 
energetic charged particles, so 
they are widely studied 
in plasma physics
\citep{Schlickeiser2002, Qin2007, 
Shalchi2009, qz2014, QinAShalchi14, 
Shalchi2020b}. 
In addition, the impacts of along-field
adiabatic focusing effect on the 
along- and cross-field diffusion 
have been extensively studied
\citep{Roelof1969, Earl1976, 
Kunstmann1979,
BeeckEA1986, BieberEA1990, 
Kota2000, SchlickeiserEA2008,
Shalchi2011, Litvinenko2012a,
Litvinenko2012b, ShalchiEA2013,
HeEA2014, WangQin2016, 
WangEA2017b, wq2018, wq2019}.

For the along-field diffusion,
three different 
definitions of diffusive
coefficient have been proposed
in the past decades, 
i.e., 
the displacement variance definition
\begin{eqnarray}
\kappa_{zz}^{DV}=\frac{1}{2}\lim_{t\rightarrow t_{\infty}}
\frac{\dee \sigma^2}{\dee t}
\end{eqnarray}
with the first- and second-order moments of charged particle 
distribution function,
the Fick's law definition
\begin{eqnarray}
\kappa_{zz}^{FL}=\frac{J}{X}
\end{eqnarray}
with $X=\partial{F}/\partial{z}$,
and the TGK formula definition
\begin{eqnarray}
\kappa_{zz}^{TGK}=\int_0^{\infty}dt
\langle v_z(t)v_z(0) \rangle.
\end{eqnarray}
If the mean magnetic field 
is uneven along the field lines, 
it has been proved that 
different definitions of 
along-field diffusion coefficient 
are not equivalent to
each other, i.e.,     
$\kappa_{zz}^{DV}$, rather than 
$\kappa_{zz}^{FL}$
and 
$\kappa_{zz}^{TGK}$, 
is the most appropriate definition
\citep{wq2018, wq2019}.
In addition, it is demonstrated 
that the cross-field diffusion coefficient 
$\kappa_\perp$ is modified 
by along-field non-uniformity of
the mean magnetic field
\citep{WangEA2017b}. Moreover,
the influence of 
along-field adiabatic focusing 
on the momentum transport 
have also been investigated
\citep{SchlickeiserEA2008,
LitvinenkoEA2011, 
WangEA2021}. 

Cross-field diffusion,
i.e., perpendicular diffusion, 
is another crucial transport
process for charged particles
in both space physics  
and laboratory plasma physics. 
Cross-field diffusion
coefficient, which is the
important parameter describing 
perpendicular transport, 
has been widely investigated in
previous studies 
\citep{MatthaeusEA2003, Shalchi2010, qz2014, 
	Shalchi2017, Shalchi2019, 
	Shalchi2020a, Shalchi2020b,
	Shalchi2021, Shalchi2021b, Shalchi2022}. 
A large number of studies have
demonstated that along-field diffusion
has a strong influence  
on perpendicular transport
\citep{MatthaeusEA2003, ShalchiEA2004, ShalchiEA2006,
	Shalchi2008, Shalchi2010, 
	Shalchi2017, Shalchi2018,
	Shalchi2019, Shalchi2020a,
	Shalchi2020b, Shalchi2021}. 
In addition, 
along-field 
adiabatic focusing effect 
is another factor influencing
cross-field diffusion
\citep{WangEA2017b}. 
However, the effect of  
solar wind on cross-field diffusion has not been 
investigated in the previous 
paper. 
In this paper, 
we also explore this problem. 
 
The remainder of this paper is 
organized as follows. 
In Section \ref{The focusing equation},
the focusing equation that satisfies 
particle number conservation is introduced. 
In Section \ref{The 
diffusion coefficients including solar wind effects},  
the along- and cross-field 
diffusion coefficients
of energetic charged particles
including solar wind effects
are derived. 
In Section \ref{The parallel diffusion coefficient for constant even magnetic field without perpendicular and 
radial momentum transport 
effects}, the influence of the
solar wind effect
on along-field diffusion 
is explored, with the dimensionless 
quantities determining the relative importance of the solar wind effect
to diffusion transport derived. 
In Section 
\ref{The mean square
displacement definition of 
perpendicular diffusion  
along x direction}, 
the effect of solar wind on cross-field diffusion
is expored. We conclude
and summarize our results in Section 
\ref{SUMMARY AND CONCLUSION}.

\section{The focusing equation}
\label{The focusing equation}

The Fokker-Planck equation is 
formulated as follows 
\citep{DuderstadtEA1979, HuangEA2008, Zank2014}
\begin{eqnarray}
\frac{\partial{f}}{\partial{t}}
+\nabla\cdot \left(\bm {u}f\right)
+\nabla_p \cdot \left(\bm {a}f\right)
=&&\nabla\cdot\left(\kappa_1\cdot \nabla 
f\right)
+\nabla_p\cdot\left(\kappa_{1}^p\cdot 
\nabla_p f\right)
+\nabla\cdot\left[\kappa_{21}\cdot 
\nabla \left(\kappa_{22}\cdot 
\nabla f\right)\right]
\nonumber\\
&&
+\nabla_p\cdot\left[\kappa_{21}^p\cdot 
\nabla_p \left(\kappa_{22}^p\cdot 
\nabla_p f\right)\right]
+\cdots.
\label{original FP equation}
\end{eqnarray}
Here, $f=f(\bm{r}, \bm{p}, t)$ 
is the distribution function
of charged energetic particles, 
$t$ is time, 
$\bm{u}$ is particle
velocity, and 
$\bm{a}$ is particle 
acceleration. 
In addition, the operators
$\nabla$ and $\nabla_p$ are
the spatial and momentum Laplacians,
respectively. 
For simplicity, 
only the first- and second-order 
derivative terms 
of the Fokker-Planck equation
are usually retained
\citep{HuangEA2008}, thus
Equation (\ref{original FP equation}) 
becomes
\begin{eqnarray}
\frac{\partial{f}}{\partial{t}}
+\nabla\cdot \left(\bm {u}f\right)
+\nabla_p \cdot \left(\bm {a}f\right)
=\nabla\cdot
\left(\kappa\cdot \nabla f\right)
+\nabla_p\cdot
\left(\kappa_p\cdot \nabla_p f\right).
\label{FP equation}
\end{eqnarray}
In general, 
the terms of the 
first- and second-order derivative
in Equation (\ref{FP equation})
can describe most of 
the important specific physical effects
of energetic particle transport
in solar wind plasmas,
e.g., 
pitch-angle scattering, 
cross-field diffusion, 
along-field adiabatic focusing,
adiabatic cooling, 
along-field spatial  
convection, etc.  
To encompasses  
all of the aforementioned 
physical processes,
the focusing
equation becomes the Fokker-Planck 
equation, 
\begin{eqnarray}
&&\frac{\partial{f}}{\partial{t}}
=\nabla\cdot\left(\kappa_\perp\cdot 
\nabla f\right)
-\nabla\cdot \left[\left(u\mu 
\hat{\bm{b}}+\bm{V}\right)f\right]
+\frac{\partial{}}{\partial{\mu}}
\left(D_{\mu\mu}\frac{\partial{f}}
{\partial{\mu}}\right)
\nonumber\\
&&
+\frac{1}{p^2}\frac{\partial{}}
{\partial{p}}
\Bigg\{p^3\Bigg[\frac{1-\mu^2}
{2}\left(\nabla\cdot 
\bm{V}-\hat{\bm{b}}\hat{\bm{b}}
:\nabla \bm{V}\right)
+\mu^2\hat{\bm{b}}
\hat{\bm{b}}:\nabla \bm{V}
\Bigg]f\Bigg\}
\nonumber\\
&&
+\frac{\partial{}}{\partial{\mu}}
\Bigg\{
\frac{1-\mu^2}{2}\Bigg[-\frac{u}{L}-
\mu\left(\nabla\cdot 
\bm{V}-3\hat{\bm{b}}\hat{\bm{b}}
:\nabla \bm{V}\right)\Bigg]f
\Bigg\},
\label{equation satisfying particle 
number conservation}
\end{eqnarray}
where
$\kappa_\perp$ is the perpendicular 
diffusion coefficient tensor,
$u$ is the particle speed,
$\mu$ is pitch-angle cosine,
$\hat{\bm{b}}$ is the unit vector
along the background magnetic field,
$p$ is the magnitude of momentum, 
$D_{\mu\mu}(\mu)$ is 
the pitch-angle diffusion 
coefficient, $L=\left(\hat{\bm{b}}\cdot\nabla\ln B\right)^{-1}$ with 
solar mean magneitc field 
$B$
is the 
characteristic length of
the adiabatic focusing,
and $\bm{V}$ is the solar wind 
velocity. 
The latter equation satisfies 
the conservation law of 
particle number, and the detailed 
derivative is shown in Appendix
\ref{The focusing equation
satisfying particle number 
conservation law}.
For convenience, the focusing 
equation can be rewritten as follows
\begin{eqnarray}
	&&\frac{\partial{f}}{\partial{t}}
	=\nabla\cdot
	\left(\kappa_\perp\cdot \nabla f\right)
	-\nabla\cdot \left[\left(u\mu \hat{\bm{b}}+\bm{V}\right)f\right]
	+\frac{\partial{}}{\partial{\mu}}
	\left(D_{\mu\mu}\frac{\partial{f}}{\partial{\mu}}\right)
 +\frac{\partial{}}{\partial{\mu}}
	\left[
	\frac{u\left(1-\mu^2\right)}{2L}f
 \right]
	\nonumber\\
	&&
	+\frac{1}{p^2}\frac{\partial{}}{\partial{p}}
	\Bigg\{p^3\Bigg[\frac{1-\mu^2}{2}
	\left(\frac{\partial{V_{x}}}
	{\partial{x}}
	+\frac{\partial{V_{y}}}
	{\partial{y}}\right)
	+\mu^2\frac{\partial{V_z}}
	{\partial{z}}
	\Bigg]f\Bigg\}
 -\frac{\partial{}}{\partial{\mu}}\Bigg[
\mu\left(\frac{\partial
		{V_{x}}}
	{\partial{x}}
	+\frac{\partial{V_{y}}}
	{\partial{y}}-2
	\frac{\partial{V_{z}}}
	{\partial{z}}\right)f
	\Bigg].
\label{equation satisfying particle 
number conservation
without tensor operation}
\end{eqnarray}
The derivation details from Equation
(\ref{equation satisfying particle 
	number conservation})
to (\ref{equation satisfying particle 
	number conservation
	without tensor operation}) 
are shown in Appendix
\ref{The focusing equation  without tensor operation}.

%%%%%%%%%%%%%%%%%%%%%%%%%%%%%%%%%%
%%%%%%%%%%%%%%%%%%%%%%%%%%%%%%%%%%

\section{The 
diffusion coefficients including solar wind effects}
\label{The 
	diffusion coefficients including solar wind effects}

In this section, we
explore the diffusion coefficients
of energetic charged particles
including solar wind effects. 

\subsection{The along-field 
diffusion coefficient including solar wind effects}
\label{The along-field 
diffusion coefficient including solar wind effects}

%The adiabatic focusing 
%can affect 
%the along-field diffusion of 
%SEPs. 
Firstly, we derive
the along-field diffusion 
coefficient formula 
of energetic particles
including  
solar wind. 

\subsubsection{The simplified
Fokker-Planck equation}
\label{The Fokker-Planck 
	equantion with wind and adiabatic focusing effects}

For simplificty,
the effect of
solar wind spatial gradient 
are ignored. 
Moreover, by performing 
the integration
$\int\dee x\int\dee y$
on Equation (\ref{equation 
	satisfying particle number 
	conservation
	without tensor operation}),
we find 
\begin{eqnarray}
	\frac{\partial{f_a}}{\partial{t}}
	=
	-u\mu\frac{\partial{f_a}}
	{\partial{z}}
	-\frac{\partial{}}
	{\partial{z}}
	\left( \int\dee x\int\dee y V_zf\right)
	+\frac{\partial{}}{\partial{\mu}}
	\left[D_{\mu \mu}\frac{\partial{f_a}}
	{\partial{\mu}}-\frac{u\left(1-\mu^2\right)}{2L}
	f_a \right].
	\label{fb equation with wind and 
		adiabatic focusing effects}
\end{eqnarray} 
Here, $f_a(z, \mu,t)=
\int \dee x\int \dee y f
(x, y, z, \mu, t)$ is
the distribution function.
%In addition, the assumption that
%$V_z$ is independent of variables
%$x$ and $y$ is used. 
Now, 
we obtain the simplified 
Fokkerr-Planck equation. 

\subsubsection{The anisotropic 
	distribution function $g_a(z,
	\mu, t)$}
\label{The anisotropic 
	distribution function gb}

The distribution function 
of charged energetic particles
can be 
divided into the isotropic part
$F_a$, which satisfies 
\begin{eqnarray}
	&&F_a=\frac{1}{2}\int_{-1}^1\dee \mu
	f_a,
	\label{F_s=int dmu f_s}
\end{eqnarray}
and the anisotropic component, which 
satisfies 
the condition
\begin{eqnarray}
	\int_{-1}^1\dee \mu g_a=0.
	\label{int dmu g_s=0}
\end{eqnarray} 
That is, the following formula holds
\begin{eqnarray}
	&&f_a=F_a+g_a,
	\label{fs=Fs+gs}
\end{eqnarray}
By integrating Equation 
(\ref{fb equation with wind and 
	adiabatic focusing effects})
over $\mu$ from $-1$ to $1$,
we can obtain
\begin{eqnarray}
	\frac{\partial{F_a}}{\partial{t}}
	=
	-\frac{u}{2}
	\frac{\partial{}}
	{\partial{z}} 
	\int_{-1}^1\dee \mu\mu g_a
	-\frac{\partial{}}{\partial{z}}
	\left(\int\dee x\int\dee y V_zF\right).
	\label{Fb equation with int gb}
\end{eqnarray} 
Here, $F=\int_{-1}^1\dee\mu 
f/2$ is employed. 
In addition, 
the boundary condition 
$D_{\mu\mu}(\mu=\pm 1)=0$ is also used. 
Similarly, integrating Equation
(\ref{fb equation with wind and 
	adiabatic focusing effects})
over $\mu$ from $-1$ to $\mu$, 
we obtain
\begin{eqnarray}
	\frac{\partial{F_a}}{\partial{t}}
	(\mu+1)
	&&+\frac{\partial{}}{\partial{t}}
	\int_{-1}^{\mu}\dee \nu g_a
	=
	-u\frac{\mu^2-1}{2}
	\frac{\partial{F_a}}
	{\partial{z}}
	-u\frac{\partial{}}
	{\partial{z}} 
	\int_{-1}^{\mu}\dee \nu
	\nu g_a
	-(\mu+1)\frac{\partial{}}{\partial{z}}
	\left(\int\dee x\int\dee y V_zF\right)
	\nonumber\\
	&&
	-\frac{\partial{}}{\partial{z}}
	\left(\int\dee x\int\dee y V_z
	\int_{-1}^{\mu}\dee\nu g(\nu)\right)
	+D_{\mu \mu}
	\frac{\partial{g_a}}{\partial{\mu}}
	-\frac{u\left(1-\mu^2\right)}{2L}
	F_a 
	-\frac{u\left(1-\mu^2\right)}{2L}
	g_a
	\label{fb equation int mu from -1 to
		mu}
\end{eqnarray}
with $g(x,y,z,\mu,t)
=f(x,y,z,\mu,t)-F(x,y,z,t)$. 
Equation
(\ref{fb equation int mu from -1 to
	mu})
can be rewritten as
\begin{eqnarray}
	&&
	\frac{\partial{g_a}}{\partial{\mu}}
	+\frac{u\left(1-\mu^2\right)g_a}
	{2D_{\mu \mu}L}
	+\frac{u\left(1-\mu^2\right)}
	{2D_{\mu \mu}}
	\left(
	\frac{\partial{F_a}}{\partial{z}}
	+\frac{F_a}{L}\right)
	=\Phi_a(\mu)
	\label{101-fb}
\end{eqnarray}
with
\begin{eqnarray}
	\Phi_a(\mu)=&&
	\frac{1}{D_{\mu\mu}}
	\Bigg[\frac{\partial{F_a}}{\partial{t}}
	(\mu+1)
	+\frac{\partial{}}{\partial{t}}
	\int_{-1}^{\mu}\dee \nu 
	g_a
	\nonumber\\
	&&
	+u\frac{\partial{}}
	{\partial{z}} 
	\int_{-1}^{\mu}\dee \nu
	\nu g_a
	+(\mu+1)\frac{\partial{}}{\partial{z}}
	\left(\int\dee x\int\dee y V_zF\right)
	+\frac{\partial{}}{\partial{z}}
	\left(\int\dee x\int\dee y V_z
	\int_{-1}^{\mu}\dee\nu g(\nu)\right)\Bigg]
	\label{Phi-fb}
\end{eqnarray}
To continue, Equation
(\ref{101-fb}) can be rewritten as
\citep{HeEA2014, wq2018, wq2019}
\begin{equation}
	\frac{\partial{}}{\partial{\mu}}
	\Bigg\{\Bigg[g_a(\mu)
	+L\left(\frac{\partial{F_a}}
	{\partial{z}}
	+\frac{F_a}{L} \right)
	\Bigg]e^{-M(\mu)}\Bigg\}
	=e^{-M(\mu)}\Phi_a(\mu)
	\label{502-fb}
\end{equation}
with
\begin{eqnarray}
	M(\mu)&=&-\frac{u}{2L}
	\int_{-1}^{\mu} d\nu
	\frac{1-\nu^2}{D_{\nu \nu}(\nu)},
	\label{M-fb}
\end{eqnarray}
By integrating
Equation (\ref{502-fb})
over $\nu$ from $-1$ to $\mu$,
we can obtain the anisotropic
distribution function
as follows
\begin{equation}
	g_a(\mu)=-L\left(\frac{\partial{F_a}}
	{\partial{z}}
	+\frac{F_a}{L}\right)\left[1-
	\frac{2e^{M(\mu)}}{\int_{-1}^{1}
		d\mu e^{M(\mu) }}\right]
	+e^{M(\mu)}\left[R_a(\mu)
	-\frac{\int_{-1}^{1}d\mu
		e^{M(\mu)}R_a(\mu)}
	{\int_{-1}^{1}d\mu
		e^{M(\mu) }}\right]
	\label{gb}
\end{equation}
with
\begin{eqnarray}
	R_a(\mu)=\int_{-1}^{\mu} d\nu
	e^{-M(\nu)}\Phi_a(\nu).
	\label{Rb}
\end{eqnarray}

\subsubsection{The governing equation
	of the isotropic distribution function
	$F_a(z, t)$}

In order to derive the 
governing equation
of the isotropic distribution function
$F_a(z, t)$, we have to
deduce the following integral
\begin{eqnarray}
	\frac{u}{2}
	\frac{\partial{}}
	{\partial{z}} 
	\int_{-1}^1\dee \mu\mu g_a
	&&
	=
	\left(
	u
	\frac{\partial{F_a}}{\partial{z}} 
	+uL
	\frac{\partial^2{F_a}}
	{\partial{z^2}}
	\right)
	\frac{\int_{-1}^1\dee \mu\mu e^{M(\mu)}}{\int_{-1}^{1}
		d\mu e^{M(\mu) }}
	\nonumber\\
	&&
	+\frac{u}{2}\int_{-1}^1\dee \mu\mu  e^{M(\mu)}\left[\frac{\partial{}}{\partial{z}} R_a(\mu)
	-\frac{1}
	{\int_{-1}^{1}d\mu
		e^{M(\mu) }}\int_{-1}^{1}d\mu
	e^{M(\mu)}\frac{\partial{}}{\partial{z}} 
	R_a(\mu)\right].
	\label{int dmu mu g}
\end{eqnarray}
The latter formula 
shows that the derivative of
$R_a(z,\mu,t)$ 
with respect to $z$ has to be
deduced
\begin{eqnarray}
	\frac{\partial{}}{\partial{z}}R_a(\mu)
	=&&
	\frac{\partial^2{F_a}}
	{\partial{z}\partial{t}}
	\int_{-1}^{\mu} d\nu
	\frac{e^{-M(\nu)}}{D_{\nu\nu}}
	(\nu+1)
	+
	\int_{-1}^{\mu} d\nu
	\frac{e^{-M(\nu)}}{D_{\nu\nu}}
	(\nu+1)
	\frac{\partial^2{}}{\partial{z^2}} 
	\left(\int\dee x\int\dee y V_zF\right)
	\nonumber\\
	&&
	+
	\int_{-1}^{\mu} d\nu
	\frac{e^{-M(\nu)}}{D_{\nu\nu}}
	\frac{\partial^2{}}
	{\partial{t}\partial{z}}
	\int_{-1}^{\nu}\dee \rho \int\dee x\int\dee y V_zF
	+u
	\int_{-1}^{\mu} d\nu
	\frac{e^{-M(\nu)}}{D_{\nu\nu}}
	\frac{\partial^2{}}{\partial{z^2}} 
	\int_{-1}^{\nu}\dee \rho
	\rho g_a
	\nonumber\\
	&&
	+
	\int_{-1}^{\mu} d\nu
	\frac{e^{-M(\nu)}}{D_{\nu\nu}}
	\frac{\partial^2{}}{\partial{z^2}} 
	\left(\int\dee x\int\dee y
	V_z \int_{-1}^{\nu}\dee \rho g_a\right).
	\label{dR/dz}
\end{eqnarray}
With Equations (\ref{int dmu mu g})
and (\ref{dR/dz}), 
the governing equation of 
the isotropic distribution function
can be found
\begin{eqnarray}
\frac{\partial{F_a}}{\partial{t}}
=&&
-\frac{\partial{}}{\partial{z}}
\left(\int\dee x\int\dee yV_z F_a\right)
-\kappa_z 
\frac{\partial{F_a}}
{\partial{z}}
+\kappa_{zz}
\frac{\partial^2{F_a}}
{\partial{z^2}}
+\kappa_{tz}
\frac{\partial^2{F_a}}
{\partial{t}\partial{z}}
+\kappa_{zz}^{V}
\frac{\partial^2{}}
{\partial{z^2}}
\left(
\int\dee x\int\dee yV_z F_a
\right)
\label{Fb equation}
\end{eqnarray}
with
and
\begin{eqnarray}
	&&\kappa_z=
	-u\frac{\int_{-1}^{1}d\mu
		\mu e^{M(\mu)}}
	{\int_{-1}^{1}
		d\mu e^{M(\mu)}},
	\label{kz}
	\\
	&&\kappa_{zz}=
	-uL
	\frac{\int_{-1}^1\dee\mu\mu e^{M(\mu)}}{\int_{-1}^{1}
		d\mu e^{M(\mu) }}
	\nonumber\\
	&&
	+
	\frac{u^2}{2}\int_{-1}^1\dee\mu\mu e^{M(\mu)}
	\int_{-1}^{\mu} d\nu
	\frac{e^{-M(\nu)}}{D_{\nu\nu}}
	\int_{-1}^{\nu}\dee\rho\rho 
	\left[1-
	\frac{2e^{M(\rho)}}{\int_{-1}^{1}
		d\mu e^{M(\mu) }}\right]
	\nonumber\\
	&&
	-
	\frac{u^2}{2} 
	\frac{\int_{-1}^1\dee\mu\mu e^{M(\mu)}}
	{\int_{-1}^{1}d\mu
		e^{M(\mu) }}\int_{-1}^{1}d\mu
	e^{M(\mu)}
	\int_{-1}^{\mu} d\nu
	\frac{e^{-M(\nu)}}{D_{\nu\nu}}
	\int_{-1}^{\nu}\dee\rho\rho 
	\left[1-
	\frac{2e^{M(\rho)}}{\int_{-1}^{1}
		d\mu e^{M(\mu) }}\right],
	\label{kzz}
	\\
&&\kappa_{tz}=
-\frac{u}{2}\int_{-1}^1\dee\mu\mu 
e^{M(\mu)}
\int_{-1}^{\mu} d\nu
e^{-M(\nu)}
\frac{\nu+1}{D_{\nu\nu}}
+\frac{u}{2} \frac{\int_{-1}^1\dee\mu\mu e^{M(\mu)}}
{\int_{-1}^{1}d\mu
	e^{M(\mu) }}\int_{-1}^{1}d\mu
e^{M(\mu)}
\int_{-1}^{\mu} d\nu
e^{-M(\nu)}
\frac{\nu+1}{D_{\nu\nu}}
\nonumber\\
&&
-
\frac{u}{2}\int_{-1}^1\dee\mu\mu 
e^{M(\mu)}
\int_{-1}^{\mu} d\nu
\frac{e^{-M(\nu)}}{D_{\nu\nu}}
\int_{-1}^{\nu}\dee\rho
\left(
\frac{2e^{M(\rho)}}{\int_{-1}^{1}
	d\mu e^{M(\mu) }}-1\right)
\nonumber\\
&&
+
\frac{u}{2} \frac{\int_{-1}^1\dee\mu\mu e^{M(\mu)}}
{\int_{-1}^{1}d\mu
	e^{M(\mu) }}\int_{-1}^{1}d\mu
e^{M(\mu)}
\int_{-1}^{\mu} d\nu
\frac{e^{-M(\nu)}}{D_{\nu\nu}}
\int_{-1}^{\nu}\dee\rho
\left(
\frac{2e^{M(\rho)}}{\int_{-1}^{1}
	d\mu e^{M(\mu) }}-1\right)
\\
	&&
	\kappa_{zz}^V=
	\frac{u}{2}
	\frac{\int_{-1}^1\dee \mu\mu 
		e^{M(\mu)}}
	{\int_{-1}^{1}d\mu
		e^{M(\mu) }}\int_{-1}^{1}d\mu
	e^{M(\mu)}
	\int_{-1}^{\mu} d\nu
	\frac{e^{-M(\nu)}}{D_{\nu\nu}}
	(\nu+1)
	\nonumber\\
	&&
	-
	\frac{u}{2}
	\int_{-1}^1\dee \mu\mu  e^{M(\mu)}
	\int_{-1}^{\mu} d\nu
	\frac{e^{-M(\nu)}}{D_{\nu\nu}}
	(\nu+1).
	\label{kzzV}
\end{eqnarray} 
Here, only the terms containing the 
first- and second-order
derivatives are retained.
In fact, the higher-order derivative 
terms do not affect the results obtained
in this article
\citep{wq2018, wq2019}.

\subsubsection{The along-field
	diffusion coefficient formula}
\label{The parallel
	diffusion coefficient formula
	for f}

To derive the formula 
of mean square displacement 
definition,
we have to obtain
the first- and second-order
moments of the isotropic 
distribution function, which
are shown as follows
\begin{eqnarray}
	&&\frac{\dee }{\dee t}
	\left\langle z\right\rangle
	=
	\left\langle V_z\right\rangle
	+\kappa_z,\\
	&&\frac{\dee }{\dee t}
	\left\langle z^2\right\rangle
	=2\left\langle zV_z \right\rangle
	+2\kappa_z\langle z \rangle
	+2\kappa_{zz}
	-2\kappa_{tz}
	\frac{\dee }{\dee t}\langle z \rangle
	+2\kappa_{zz}^V 
	\left\langle V_z\right\rangle.
\end{eqnarray}
Combining the latter formulas gives
\begin{eqnarray}
	\frac{1}{2}\frac{\dee \sigma^2}{\dee t}
	=\frac{1}{2}\frac{\dee }{\dee t}
	\left(\left\langle z^2\right\rangle
	-\langle z \rangle^2\right)
	=T_1-T_2+T_3+T_4
	\label{dsigma2/2dt-Fb}
\end{eqnarray}
with
\begin{eqnarray}
	&&T_1=\kappa_{zz},\\
	&&T_2=\kappa_{tz}\kappa_z,\\
	&&T_3=\zeta=\left\langle zV_z \right\rangle
	-\langle z \rangle
	\left\langle V_z\right\rangle,
	\label{T3}\\
	&&T_4=\left(\kappa_{zz}^V 
	-
	\kappa_{tz}\right)
	\left\langle V_z\right\rangle.
	\label{T4}
\end{eqnarray}
From Equations 
(\ref{T3}) and (\ref{T4}),
we can find that
the mean square displacement 
definition of
along-field diffusion coefficient
includes the solar wind
and adiabatic focusing effects. 
According to the results
obtained by \citet{wq2019},  
the term $T_4$ is approximately 
equal to zero. Thus, 
Equation (\ref{dsigma2/2dt-Fb}) 
becomes
\begin{eqnarray}
	\frac{1}{2}\frac{\dee \sigma^2}{\dee t}
	=\frac{1}{2}\frac{\dee }{\dee t}
	\left(\left\langle z^2\right\rangle
	-\langle z \rangle^2\right)
	=T_1-T_2+T_3.
	\label{dsigma2/2dt-Fb-2}
\end{eqnarray}

Note that if we 
only consider adiabatic 
focusing, Equation (\ref{dsigma2/2dt-Fb-2})
becomes
\begin{eqnarray}
	\frac{1}{2}\frac{\dee \sigma^2}{\dee t}
	=T_1-T_2
	=\kappa_{zz}
	-\kappa_{tz}\kappa_z,
\end{eqnarray}
which is identical to the result derived
by \citet{wq2018}. 

\subsection{The cross-field 
	diffusion coefficient including solar wind effects}
\label{The cross-field 
	diffusion coefficient including solar wind effects}

Next, we derive the cross-field
diffusion coefficient of energetic
charged particles including
solar wind effect.

\subsubsection{The govering equation
	of isotropic distribution function}

The starting point of the investigation
in this subsection is also the focusing
equation, which is displayed in 
Section \ref{The focusing equation}. 
For Equation 
(\ref{equation satisfying particle number conservation
without tensor operation}),
by ignoring the terms containing spatial
derivaitve of solar wind speed
and integrating 
over $\mu$ from $-1$ to $1$,
we can obtain
the governing equation 
of isotropic distribution
function, which is given as follows
\begin{eqnarray}
	\frac{\partial{F'}}{\partial{t}}
	=\nabla\cdot
	\left(\kappa_\perp\cdot 
	\nabla F'\right)	
	-\frac{u}{2}\frac{\partial{}}
	{\partial{z}}
	\int_{-1}^{1}\dee\mu\mu g
	-\nabla\cdot 
	\left(\bm{V} F'\right)
	\label{F' equation with g}
\end{eqnarray}
with
\begin{eqnarray}
	F'=\frac{1}{2}
	\int_{-1}^{1}\dee\mu f.
\end{eqnarray}
Here, $g(x,y,z,\mu,t)$ is the anisotropic 
distribution function. 
Performing integrating 
$\int \dee y\int \dee z$ on
Equation (\ref{F' equation with g})
yields
\begin{eqnarray}
	\frac{\partial{F_b}}{\partial{t}}
	=\frac{\partial{}}{\partial{x}}
	\left(\kappa_\perp\frac{\partial{F_b}}
	{\partial{x}}\right)
	-\frac{\partial{}}{\partial{x}} \left(\int \dee y\int \dee zV_x F' \right).
	\label{Fc equation}
\end{eqnarray} 
Here, $F_b=F(x, t)$ is the isotropic distribution function satisfying
the following formula
\begin{eqnarray}
	F_b=\frac{1}{2}
	\int_{-1}^{1}\dee\mu
	\int \dee y\int \dee z f. 
\end{eqnarray}

\subsubsection{The mean square
	displacement definition of 
	perpendicular diffusion  
in $x$ direction}

From Equation 
(\ref{Fc equation}), the first-
and second-order moments of
charged particle distribution 
function can be obtained as
\begin{eqnarray}
	&&\frac{\dee}{\dee t}\langle x \rangle
	=\int \dee x \int \dee y \int \dee z
	V_x F'
	=\left\langle V_x \right\rangle,\\ 
	&&\frac{\dee}{\dee t}
	\left\langle x^2 \right\rangle
	=2\kappa_\perp
	+2\int \dee x \int \dee y \int \dee z xV_x F’
	=2\kappa_\perp
	+2\left\langle xV_x \right\rangle.
\end{eqnarray}
Combining the latter formulas gives
\begin{eqnarray}
	\kappa_{xx}^{DV}=
	\frac{1}{2}\frac{\dee \sigma_x^2}
	{\dee t}
	=\frac{1}{2}\frac{d}{dt}
	\left(\langle x^2 \rangle
	-\langle x\rangle^2\right)
	=\kappa_\perp
	+\eta
	\label{mean square displacement definition of kxx}  
\end{eqnarray}
with
\begin{eqnarray}
	\eta=\left\langle xV_x \right\rangle
	-\langle x \rangle
	\left\langle V_x \right\rangle.
	\label{B}	
\end{eqnarray}
Here, $\kappa_\perp=\kappa_{xx}^{FL}$
is the Fick's law definition. 
In addition,
$\eta$ is the solar wind
effect on the cross-field
diffusion. 
Thus,
Equation 
(\ref{B}) indicates that
the cross-field
diffusion is affected by
solar wind effect.  

In the following, we would 
discuss the influence of
solar wind on along- and cross-field diffusion, 
i.e., Equations 
(\ref{T3}) and (\ref{B}), 
with the typical parameter
values, e.g., 
solar wind 
speed $V\sim3
\times10^5$ m/s, 
the solar rotation speed
$\omega\sim 3\times10^{-6}$ /s,
the energetic proton speed 
$u\sim 10^{7}$ m/s, 
the particle parallel
mean free path $\lambda\sim
10^{10}$ m, 
and the parameter 
$\lambda_\perp \equiv 3\kappa_\perp/u
=10^9$ m. 

\section{The influence of
solar wind on along-field diffusion}
\label{The parallel diffusion coefficient for constant even magnetic field without perpendicular and 
radial momentum transport effects}

In this section,
based on Equation 
(\ref{dsigma2/2dt-Fb-2}) 
we discuss the 
solar wind effect
on along-field diffusion
of energetic charged particles.
If we only consider 
solar wind effect, 
Equation (\ref{dsigma2/2dt-Fb-2}) is
simplifies as
\begin{eqnarray}
\kappa_{zz}^{DV}=
\frac{1}{2}\frac{\dee \sigma^2}{\dee t}
	=T_1+T_3
	=\kappa_{zz}+\zeta,
\label{dsigma/dt=kzz+zeta}
\end{eqnarray}
with
\begin{eqnarray}
	\zeta=\left\langle zV_z\right\rangle
	-\left\langle z\right\rangle
	\left\langle V_z\right\rangle.
	\label{A}
\end{eqnarray} 
Here, 
$\sigma^2=
\left\langle
z^2\right\rangle
-\left\langle
z\right\rangle^2$,
and $\zeta$ is the influence of
solar wind on along-field
diffusion coefficient.  
It is obvious that Equation 
(\ref{dsigma2/2dt-Fb}) includes 
not only adiabatic focusing effect
but also solar wind effect. 
Equations (\ref{dsigma/dt=kzz+zeta})
and (\ref{A}) show
that
the mean square displacement 
definition $\kappa_{zz}^{DV}$ 
of energetic charged particles
is not equal to
the Fick' law definition 
$\kappa_{zz}^{FL}$. 
According to the results found by
\citet{wq2019}, with along-field 
adiabatic focusing,
the mean square displacement 
definition $\kappa_{zz}^{DV}$ 
is more appropriate than both
Fick' law one
$\kappa_{zz}^{FL}$
and the Taylor-Green-Kubo one
$\kappa_{zz}^{TGK}$. 
If the same operation 
performed as 
\citet{wq2019} for Equation
(\ref{Fb equation}),  
the similar 
conclusion can be easily
obtained. 

Equations 
(\ref{dsigma/dt=kzz+zeta}) and (\ref{A}) show that 
the mean square displacement 
definition $\kappa_{zz}^{DV}$
accounts for the solar wind
effect.   
Therefore, when the transport
of SEPs
parallel to the mean magnetic field
is investigated, 
the relative importance of 
the solar wind effect to 
along-field diffusion  
has to be explored. 

\subsection{The integrating form 
of the solar wind 
effect, $\zeta$}
\label{The integrating form 
of stellar wind effect A}

\begin{figure}
\centering
\includegraphics[width=0.7\textwidth]
{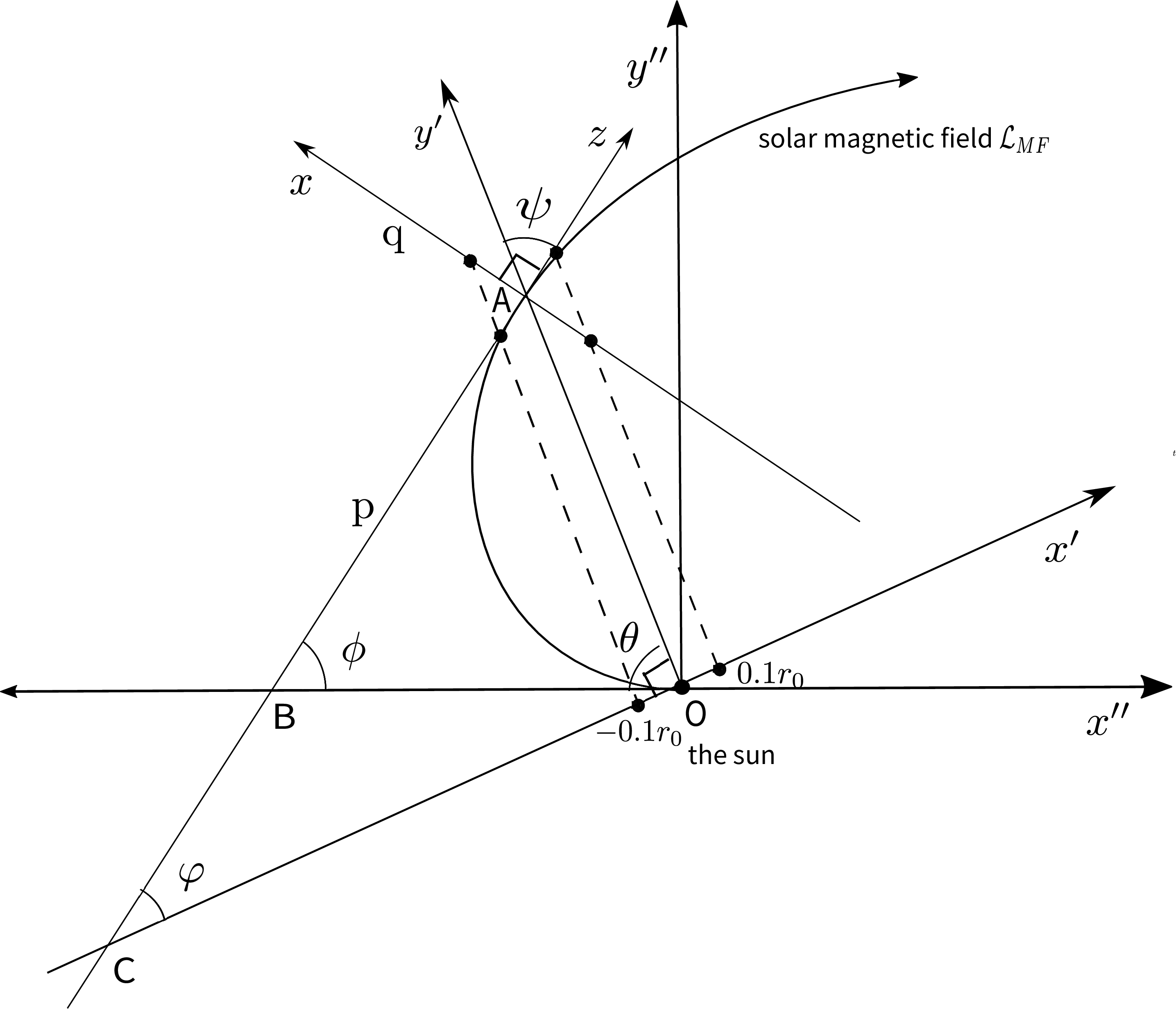}
\caption{
Schematic diagram of
magnetic coordinate system $x-y-z$, 
solar coordinate systems
$x'-y’-z’$ and $x''-y''-z''$, 
the polar coordinate system,
which are in ecliptic plane,
and $z' \parallel z''$. 
}

\label{fig1}
\end{figure}

In order to investigate the relative
importance of the solar wind
effect on along-field diffusion,
it is necessary to derive
the specific form of the  
solar wind effect, 
$\zeta$. 
Due to the magnetic freezing effect 
of plasmas and the solar rotation, 
the mean magnetic field of the sun
presents a spiral form, which is called
the Parker spiral.
For simplicity, we consider it  in the 
ecliptic plane. 
As shown in Figure 
\ref{fig1},  
we consider a point $A$  with the 
solar magnetic field line $\mathcal{L}_{\txt{MF}}$ going through it. 
We set a coordinate system 
$x'-y'-z'$, 
with the $y'$-axis pointing 
from the sun to 
point $A$, the $z'$-axis 
perpendicular to
the ecliptic plane pointing towards
the north, 
the $x'$-axis defined using the right-hand rule,
and the origin $O$ located at 
the center of the sun. 
We also set a polar coordinate system $r-\theta$,
with the polar axis $r$ tangent to $\mathcal{L}_{\txt{MF}}$
at point $O$, 
the polar angle $\theta$ 
defined in a clockwise direction. 
In addition, another
coordinate system $x''-y’'-z''$
is established by rotating 
$x'-y'-z'$ system
with $\pi/2-\theta_A$ 
clockwise through point $O$,
where $\theta_A$ is the polar angle of point $A$. 
Additionly, we set a magnetic 
coordinate system $x-y-z$,
with the $z$-axis along 
the tangent of 
magnetic field line 
towards its positive direction,
the $y$-axis parallel to $z'$,
and the $x$-axis satisfying
right-hand rule.
Moreover, the tangent
of $\mathcal{L}_{\txt{MF}}$
at point $A$ is
the straight line
$y'=kx'+r_0$ with the intercept
$r_0$, i.e., the distance from $O$ to $A$.
We suppose the straight line
can be written as $y''=k'x''+l$
in $x''-y''-z''$ system.  
Furthermore, 
normal equation 
of $\mathcal{L}_{\txt{MF}}$
at point $A$ 
is $y'=-x'/k+r_0$. 

Additionly, 
the angle between $z$-axis
and $y'$-axis is denoted as $\psi$. 
Therefore,
the component of 
solar wind speed along 
solar magnetic field can 
be expressed as 
\begin{eqnarray}
V_z=V\cos\psi,
\end{eqnarray}    
where $V$ is the magnitude
of the solar wind velocity
$\bm{V}$. 
It is obvious that
the angle $\psi$ varies depending on
the point on the magnetic field, and 
obeys the following formula
on the ecliptic plane 
\citep[e.g.,][]{QiWu2018}
\begin{eqnarray}
\cos\psi=\left(
1+\frac{\omega^2r^2
}{V^2}\right)^{-1/2}.
\label{cos psi}
\end{eqnarray}
Here, $\omega$ is the 
angular velocity of the sun, and
$r$ is the radial distance from
the center of the sun 
to the point on 
solar magentic field. 
Thus, 
the influence $\zeta$ of 
the solar wind 
can be written as 
\begin{eqnarray}
\zeta=\int_{z_1}^{z_2}\dee zF(z, t)V
\left(
1+\frac{\omega^2r^2
}{V^2}\right)^{-1/2}
-\int_{z_1}^{z_2} \dee zz 
F(z, t)	
\int_{z_1}^{z_2} \dee z F(z, t)V
\left(
1+\frac{\omega^2r^2
}{V^2}\right)^{-1/2},
\label{<zV>-<z><V>}
\end{eqnarray}
where $z_1$ and $z_2$ are the 
integral lower and upper limits,
respectively. 
The integrals in Equation
(\ref{<zV>-<z><V>})
are along the curve of the 
mean magnetic field,
and
the normalization condition
needs to be satisfied
\begin{eqnarray}
\int_{z_1}^{z_2} \dee F(z, t)=1. 
\label{normalization condition 
from z1 to z2}
\end{eqnarray}  

\subsection{Exploring the relative
importance of solar
wind effect to along-field 
diffusion}
\label{Exploring the 
importance of background plasma 
wind effect in the parallel diffusion}

In this paper, for the sake of
simplicity, 
we  assume that  
the speed $V$ of the solar wind 
is constant. Thus, Equation
(\ref{<zV>-<z><V>})
can be written as
\begin{eqnarray}
\zeta=V\int_{z_1}^{z_2}\dee zz F(z, t)
\left(
1+\frac{\omega^2r^2
}{V^2}\right)^{-1/2}
-V\int_{z_1}^{z_2} \dee zz
F(z, t)	
\int_{z_1}^{z_2} \dee z  F(z, t)
\left(
1+\frac{\omega^2r^2
}{V^2}\right)^{-1/2}.
\label{<zV>-<z><V>-fs-0}
\end{eqnarray}
In coodinate system $x'-y'-z'$, 
if integral interval is small 
enought, 
the straight line
$y'=kx'+r_0$ with $k=\tan\beta
=\cot\psi$ can be 
used to approxmiately replace 
the curve of magnetic field line for the integrals in Equation
(\ref{<zV>-<z><V>-fs-0}), 
and the distribution
function $F(x',t)$ can be used
to replace $F(z,t)$. 
Accordingly, as shown in Figure
\ref{fig1}, 
the distance from $O$ to any point,
$(x', y')$,
between the integral interval
$[z_1, z_2]$
can be expressed as
$r=\sqrt{x'^2+(kx'+r_0)^2}$.
In addition, 
we can obtain the formulas 
$\dee z=\sqrt{1+(\dee y'/\dee x')^2}\dee x'$
and $z\approx\sqrt{1
+(\dee y'/\dee x')^2} x'$.
To proceed, if the integral
interval $[z_1, z_2]$
is set as 
$[-0.1r_0, 0.1r_0]$ 
in coordinate system
$x'-y'-z'$, 
with the above setting,
Equation 
(\ref{<zV>-<z><V>-fs}) becomes
\begin{eqnarray}
\zeta=&&V\left(1+k^2\right)
\int_{-0.1r_0}^{0.1r_0}
\dee x'x' F(x', t)
\left(1+\frac{\omega^2
\left[x'^2+(kx'+r_0)
^2\right]}{V^2}
\right)^{-\frac{1}{2}}
\nonumber\\
&&
-V\left(1+k^2\right)^{3/2}
\int_{-0.1r_0}^{0.1r_0} \dee x'x' 
F(x', t)	
\int_{-0.1r_0}^{0.1r_0} 
\dee x' F(x', t)
\left(1+\frac{\omega^2
\left[x'^2+(kx'+r_0)
^2\right]}{V^2}
\right)^{-\frac{1}{2}}.
\label{<zV>-<z><V>-fs}
\end{eqnarray}
For mathematical tractability,
in this article, we only explore 
the tail of SEPs, for which 
the distribution function 
$F(x', t)$
is approximately 
uniform and is also approximately
an even function
of variable $x'$ in integral interval
$[-0.1r_0, 0.1r_0]$.
Thus,  
the second term 
on the right-hand side
of Equation (\ref{<zV>-<z><V>-fs})
is equal to zero, and we have 
\begin{eqnarray}
\zeta=&&V\left(1+k^2\right)
\int_{-0.1r_0}^{0.1r_0}
\dee x'x' F(x', t)
\left(1+\frac{\omega^2
	\left[x'^2+(kx'+r_0)
	^2\right]}{V^2}\right)^{-\frac{1}{2}}.
\label{<zV>-<z><V>-fs-1}
\end{eqnarray}
In fact,
based on the characterisitcs
of each SEP events,  
the integral interval can  
also be set to other 
values. In this paper, we only 
qualitatively explore 
the solar wind effect on 
energetic particle
along-field diffusion. Therefore,
the interval length in Equation
(\ref{<zV>-<z><V>-fs-1}) 
does not affect the findings
obtained in this paper. 

With the integral interval
$[-0.1r_0, 0.1r_0]$
and Equation
(\ref{normalization condition 
from z1 to z2}),
the normalization condition 
becomes
\begin{eqnarray}
	F(x',t)=\frac{5}{r_0}.
	\label{F=5/r0}
\end{eqnarray}
Accordingly,  
the background 
solar wind effect on along-field
diffusion can be written as
\begin{eqnarray}
\zeta=\left(1+k^2\right)
\frac{5V}{r_0}
\int_{-0.1r_0}^{0.1r_0}
\dee x'x'
\left(1+\frac{\omega^2
\left[x'^2+(kx'+r_0)
^2\right]}{V^2}\right)^{-\frac{1}{2}}.
\label{<zV> with x}
\end{eqnarray} 
With integration by parts and
a lengthy mathematical 
performance,
the latter formula becomes
\begin{eqnarray}
\zeta=&&
\frac{2k\frac{V^2}{\omega}}
{\sqrt{\left(0.1+\frac{1}{1+k^2}k
\right)^2\left(1+k^2\right)
+\frac{1}{\left(1+k^2\right)}
+\frac{V^2}{\omega^2r_0^2}}
+
\sqrt{\left(-0.1
+\frac{1}{1+k^2}k\right)^2
\left(1+k^2\right)
+\frac{1}{\left(1+k^2\right)}
+\frac{V^2}{\omega^2r_0^2}}}
\nonumber\\
&&
-\frac{5k}{\sqrt{1+k^2}}
\frac{V^2}{\omega}
\ln\frac{\left(0.1
+\frac{1}{1+k^2}k\right)
\sqrt{1+k^2}
+\sqrt{\left(0.1
+\frac{1}{1+k^2}k\right)^2
\left(1+k^2\right)
+\frac{1}{\left(1+k^2\right)}
+\frac{V^2}{\omega^2r_0^2}}}
{\left(-0.1+\frac{1}{1+k^2}k\right)
\sqrt{1+k^2}
+
\sqrt{\left(-0.1
+\frac{1}{1+k^2}k
\right)^2\left(1+k^2\right)
+\frac{1}{\left(1+k^2\right)}
+\frac{V^2}{\omega^2r_0^2}}}
\label{<zV>-fs after integration}.
\end{eqnarray}
Obviously, the latter equation
contains
the following two 
dimensionless quantities
\begin{eqnarray}
&&\alpha_1\equiv
\frac{V}{\omega r_0},
\label{alpha1}\\
&&\alpha_2\equiv|k|.
\label{k}
\end{eqnarray} 
For different limits of 
$\alpha_1$ and $\alpha_2$,
the relative importance of 
the solar wind effect 
on along-field diffusion, $\zeta$,
can be discussed.  
The results are summarized
in Table \ref{The cases for parallel diffusion with plasma wind effect}. 

\subsubsection{The condition
$\alpha_1\gg \alpha_2\gg 1$}
\label{case 1 for A to kzz}

Here, we suppose 
\begin{eqnarray}
\alpha_1\gg \alpha_2\gg 1,
\label{condition 1}
\end{eqnarray}
which contains 
the inequalities
\begin{eqnarray}
&&\alpha_1\gg \alpha_2,
\label{condition 11} \\
&&\alpha_1\gg 1,
\label{condition 12}\\
&&\alpha_2\gg 1.
\label{condition 13}
\end{eqnarray}
For the condition
(\ref{condition 1}),
Equation (\ref{<zV>-fs after integration})
is simplified as
\begin{eqnarray}
\zeta=|k|Vr_0.
\label{<zV>-<z><V>-fs
for alpha1 gg 1}
\end{eqnarray}
It is known that particle 
diffusion coefficient 
$\kappa_{zz}$ can be written as
\begin{eqnarray}
\kappa_{zz}=\frac{u\lambda}{3},
\label{kzz=u lambda/3}
\end{eqnarray}
where $\lambda$ is the 
mean free path of particles.
With Equations 
(\ref{<zV>-<z><V>-fs
for alpha1 gg 1}) 
and (\ref{kzz=u lambda/3}),
the relative importance
of solar wind effect, $\zeta$, 
on 
the along-field
diffusion is shown as follows 
\begin{eqnarray}
\frac{\zeta}{\kappa_{zz}}=
3|k|\frac{Vr_0}{u\lambda}.
\end{eqnarray}
It is obvious that the 
following dimensionless quantity
\begin{eqnarray}
\beta_1=|k|\frac{Vr_0}{u\lambda}
\end{eqnarray}
determines the relative importance
of $\zeta$ to $\kappa_{zz}$. 
If $\beta_1\gtrsim1$, the 
influence of  
solar wind effect, $\zeta$,
on the along-field diffusion 
should be taken into account,
conversely, if $\beta_1\ll 1$,
$\zeta$ can be
ignored.
For the typical
values of $V$, $u$, and
$\lambda$ in Section
\ref{The diffusion coefficients including solar wind effects},
the dimensionless 
quantity becomes
\begin{eqnarray}
\beta_1=|k|r_0\frac{V}{u\lambda}
\sim 10|k|r_0.
\label{case 1 equation 1}
\end{eqnarray}

Thus, we only need to 
explore the value of $kr_0$. 
The solar magnetic field
can be described by
\begin{eqnarray}
r_0=\frac{V}{\omega}\theta
\label{solar magnetic field
polar equation}
\end{eqnarray}
with solar wind speed
$V$ and solar rotation speed
$\omega$. 
It is known that 
the angle $\theta$ of polar system
inscreases in an
clockwise direction.
The parametric formulas
of Equation (\ref{solar magnetic field
	polar equation})
in coordinate system $x''-y''-z''$
are shown as 
\begin{eqnarray}
&&x''=\frac{V}{\omega}
\theta\cos\theta,\\
&&y''=\frac{V}{\omega}
\theta\sin\theta. 
\end{eqnarray}	
The slope of the tangent line 
at the point $A$ 
is $k'=\tan\phi$ with 
the angle $\phi$
between $x''$-axis and
the tangent. 
With the latter equations, 
the tangent slope of
point $A$ on the magnetic field
line in coordinate system
$x''-y''-z''$
can be written as
\begin{eqnarray} 
k'=\tan\phi
=\frac{\dee y''}{\dee x''}
=\frac{\dee y''/\dee\theta}
{\dee x''/\dee\theta}
=\frac{\theta+\tan\theta}
{1-\theta\tan\theta}.
\label{k'} 
\end{eqnarray}
Similarly, the slope 
$k$ of the tangent is 
$k=\tan\varphi$ in coordinate
system $x'-y'-z'$, which 
satisfies 
\begin{eqnarray} 
k=\tan\varphi=\tan 
\left(\phi+\theta
	-\frac{\pi}{2}\right)
	=\frac{2\theta\tan\theta+
		\tan^2\theta-1}{\theta+2\tan\theta-\theta\tan^2\theta}.
	\label{k}
\end{eqnarray}
It is easily proved that 
slope $k$ is the generally 
increasing function of variable
$\theta$. It is suggested that
the points with
$\theta=n\pi+\pi/2$ for
integer numbers $n$ 
in Equation (\ref{k}) are 
removable singularities,
which
do not affect the monotonicity
of function $k$ with $\theta$.  
Therefore, the quantity $k$ is 
an 
appropriate parameter to
reflect the influence of solar 
wind on particle transport, $\zeta$, 
at different spatial locations 
within the heliosphere. 
Accordingly, 
Equation (\ref{k}) becomes
\begin{eqnarray} 
|k|r_0=\frac{V}{\omega}\theta\left|
\frac{2\theta\tan\theta
+\tan^2\theta-1}{\theta
+2\tan\theta-\theta\tan^2
\theta}\right|.
\label{kr}
\end{eqnarray}
If $\theta\to 0$, 
for typical parameter values
$V$ and $\omega$
listed above, 
we can obtain
\begin{eqnarray}
|k|r_0\to 10^{10} \txt{m}.
\label{kr-2}
\end{eqnarray}
Therefore,
with Equation
(\ref{case 1 equation 1}),
we can find
\begin{eqnarray}
\beta_1\sim 10|k|r_0\sim 10^{11}\gg1.
\label{beta1 value for A/kzz for
inner heliosphere}
\end{eqnarray}
In addition, Inequality
(\ref{condition 12})
denotes 
$r_0\ll V/\omega\sim 10^{11} \txt{m}\sim
1$ AU,
which indicates that the point $A$
is located in the inner 
heliosphere
and Inequality 
(\ref{condition 13}). 
It is noted
that Inequality 
(\ref{condition 13}),
is consistent with the condition,
Inequality (\ref{condition 1}).
However, Inequality 
 (\ref{condition 1})
leads to the relation 
$\alpha_1/\alpha_2\gg 1$,
i.e.,
$|k|r_0\ll V/\omega$, which 
is contradictory to Equation
(\ref{kr}) in the 
inner heliosphere.
Therefore, for this case,
the influence of solar wind 
on the along-field diffusion,
$\zeta$,
may not be taken into account. 

\subsubsection{The condition
$\alpha_2\gg \alpha_1\gg 1$}
\label{case 2 for A to kzz}

In this subsection, we suppose 
\begin{eqnarray}
\alpha_2\gg \alpha_1\gg 1,
\label{condition 2}
\end{eqnarray} 
which contains
\begin{eqnarray}
&&\alpha_2\gg \alpha_1,
\label{condition 21}\\
&&\alpha_2\gg 1,
\label{condition 22}\\
&&\alpha_1\gg 1.
\label{condition 23}
\end{eqnarray}

Equation (\ref{<zV>-fs after integration}) becomes
\begin{eqnarray}
\zeta=-10\frac{V^2}{\omega}
\ln \frac{\alpha_2}{\alpha_1}.
\end{eqnarray}
With the along-field diffusion 
coefficient formula, i.e.,
Equation (\ref{kzz=u lambda/3}), 
we find
\begin{eqnarray}
\frac{\zeta}{\kappa_{zz}}=
-30\frac{V^2}{u\omega\lambda}
\ln \frac{\alpha_2}{\alpha_1}. 	
\label{A/kzz}
\end{eqnarray}
Obviously, the latter 
equation is 
determined by the following 
two dimensionless quantities
\begin{eqnarray}
&&\beta_2=\frac{V^2}
{u\omega\lambda},\\
&&\beta_3=\frac{\alpha_2}
{\alpha_1}=
\frac{\omega r_0|k|}{V}.
\end{eqnarray}
As shown in subsection 
\ref{case 1 for A to kzz},
the condition 
$\alpha_1\gg 1$ and 
$\alpha_2\gg 1$
can only be satisfied 
in the inner
heliosphere. 
However, from Inequality
(\ref{condition 21})
we can obtain 
$\alpha_2/\alpha_1=\beta_3\gg1$.
For typical parameter values
$V$ and $\omega$ in subsection
\ref{case 1 for A to kzz}, 
we find  
\begin{eqnarray}
|k|r_0\gg\frac{V}{\omega},
\label{case 2-1}
\end{eqnarray}
which cannot be satisfied in the 
inner heliosphere. 
Therefore, in this condition,
the solar wind effect on 
along-field diffusion
of energetic particles, $\zeta$,
might be ignored. 

\subsubsection{The condition
$\alpha_1\sim \alpha_2\gg 1$}
\label{alpha1=alpha2 gg 1}

In the following, we suppose 
\begin{eqnarray}
\alpha_1\sim \alpha_2\gg 1, 
\label{condition 3}
\end{eqnarray}
which contains two conditions
\begin{eqnarray}
&&\alpha_1\sim \alpha_2,\\
\label{condition 31}
&&\alpha_1\gg 1,
\label{condition 32}\\
&&\alpha_2\gg 1.
\label{condition 33}
\end{eqnarray}
The condition (\ref{condition 31})
denotes $|k|r_0\sim V/\omega$,
which is consistent with
Equation (\ref{kr}) 
at least in the inner heliosphere.
In addition, Inequality
(\ref{condition 32})
corresponds to
$r_0\ll V/\omega\sim 10^{11}m$, which 
indicates the point $A$
is located in the inner heliosphere.
Inequality (\ref{condition 33})
also shows that the point $A$
is close to the sun.  

Additionly, for the condition 
(\ref{condition 3}),
Equation (\ref{<zV>-fs after integration}) is simplified as
\begin{eqnarray}
\zeta=0.4\frac{V^2}{\omega}. 
\end{eqnarray}
The ratio of $A$ to $\kappa_{zz}$
is
\begin{eqnarray}
	\frac{\zeta}{\kappa_{zz}}
	=1.2\frac{V^2}{u\omega\lambda},
\end{eqnarray}
which shows that the following 
dimensionless quantity
determines the relative importance
of solar wind effect
\begin{eqnarray}
\beta_2=\frac{V^2}{u\omega\lambda}.
\end{eqnarray}
When $\beta_2 \sim 1$, the solar wind 
effect on along-field diffusion
of energetic particles,
$\zeta$, is important.
For the typical parameter values
$V$, $\omega$, $u$, and $\lambda$
listed in subsection 
\ref{case 1 for A to kzz}, 
we can obtain $\beta_2=0.3$.
In summary, for Inequality
(\ref{condition 3}),
i.e., the point $A$ 
located in the inner 
heliosphere, solar wind effect
on along-field diffusion,
$\zeta$, 
is relatively important,
so it should be considered.    

\subsubsection{The condition
$\alpha_1\gg1$ and $\alpha_2\ll 1$}

Here, we suppose
\begin{eqnarray}
\alpha_1\gg1,
\label{condition 41}
\end{eqnarray}
and 
\begin{eqnarray}
\alpha_2\ll 1. 
\label{condition 42}
\end{eqnarray}
For Inequalities
(\ref{condition 41}) and 
(\ref{condition 42}),
Equation 
(\ref{<zV>-fs after integration})
is simplified as
\begin{eqnarray}
\zeta=|k|Vr_0,
\end{eqnarray}
which is identical to Equation
(\ref{<zV>-<z><V>-fs
for alpha1 gg 1}). Accordingly,
the dimensionless quantity
controlling the relative
importance is 
\begin{eqnarray}
\beta_1=|k|\frac{Vr_0}{u\lambda}. 
\end{eqnarray}

In addition, 
Inequality (\ref{condition 41}), 
i.e., 
$r_0\ll V/\omega$, corresponds to
$r_0\ll 10^{11}m$,
which denotes that the point 
$A$ is in the inner heliosphere. 
However, Inequality 
(\ref{condition 42})
represents that the slope
of tangent line is very small,
which indicates that the  
point $A$ is located in the outer 
heliosphere. 
Thus,  
the two Inequalities
(\ref{condition 41}) and
(\ref{condition 42})
are mutually 
contraditory. Therefore, 
for this case,
the solar wind 
effect, $\zeta$, 
might not be considered. 

\subsubsection{The condition
$\alpha_2\gg 1$ and 
$\alpha_1\ll 1$}

In this subsection, we suppose    the two inequalities 
\begin{eqnarray}
&&\alpha_2\gg 1,
\label{condition 51}\\
&&\alpha_1\ll 1. 
\label{condition 52}	
\end{eqnarray}
Using the the latter two inequalities, 
from Equation 
(\ref{<zV>-fs after integration})
we can obtain
\begin{eqnarray}
\zeta=10\frac{V^2}{\omega}. 	
\end{eqnarray}
Comparing the latter formula with
$\kappa_{zz}$, we can find
the following dimensionless 
quantities 
\begin{eqnarray}
\beta_2=\frac{V^2}{u\omega\lambda}.	
\end{eqnarray}
Inequality (\ref{condition 51})
indicates
the point $A$ is in the inner 
heliosphere.
Additionly, 
for the typical parameter
values listed 
in Section 
\ref{The 
	diffusion coefficients including solar wind effects}, 
Inequality (\ref{condition 52})
denotes
$r_0\gg 10^{11}$ m $\sim 1$ AU
which indicates that the 
point $A$ is located in the outer heliosphere. Thus, Inequatity
(\ref{condition 51}) is contradictory
with Inequality 
(\ref{condition 52}). 
Therefore, 
for this case the solar wind 
effect, $\zeta$, on energetic particle
transport may not be 
considered. 
 
\subsubsection{The condition 
$\alpha_1\ll 1$ and 
$\alpha_2\ll 1$}

Now, we suppose
the two inequalities
\begin{eqnarray}
&&\alpha_1\ll 1,
\label{condition 61}\\
&&\alpha_2\ll 1.
\label{condition 62}
\end{eqnarray}
For the typical 
parameter values listed 
in Subsection 
\ref{case 1 for A to kzz}, 
Inequality 
(\ref{condition 61})
gives $r_0\gg10^{11}m$,
which indicates the point $A$
located in the outer heliosphere.
This is consistent with
Inequality (\ref{condition 62}).  

In addition, 
for the two Inequalitites
(\ref{condition 61})
and
(\ref{condition 62}), 
Equation (\ref{<zV>-fs after integration}) becomes
\begin{eqnarray}
\zeta=|k|\frac{V^2}{\omega}.
\label{zeta for alpha1 and 
alpha2 all ll 1}
\end{eqnarray}
By comparing the latter formula 
with $\kappa_{zz}$,
we obtain
\begin{eqnarray}
\frac{\zeta}{\kappa_{zz}}=
3\frac{|k|V^2}{u\omega\lambda},
\end{eqnarray}	
which is determined by the dimensionless 
quantities
\begin{eqnarray}
\beta_4=\frac{|k|V^2}
{u\omega\lambda}.
\label{beta3}
\end{eqnarray}
For the parameter values 
$V$, $u$, $\omega$, and $\lambda$
listed 
in Section 
\ref{The 
	diffusion coefficients including solar wind effects}, 
from 
Equation (\ref{beta3}) we can 
obtain $|k|\ge 3$, which
is contradictory to  
Inequality (\ref{condition 62}).
Therefore, when 
energetic particle transport
in the outer helioshere,
the solar wind effect, $\zeta$,
might not 
be 
taken into account.   

\subsection{Relative
importance of 
the solar wind and
adiabatic focusing effects}
\label{Exploring the influence of adiabatic focusing effect on parallel diffusion}

The second term on the right-hand 
side 
of Equation (\ref{dsigma2/2dt-Fb})
was evaluated by \citet{wq2018, wq2019} as
\begin{eqnarray}
T_2=\kappa_z\kappa_{tz}\approx 
\frac{14}{135}
\frac{u\lambda^3}{L^2}.
\label{kzktz}
\end{eqnarray}
Acccordingly, 
the ratio of adiabatic 
focusing effect term 
$T_2$
to along-field Fick's law diffusion term
$T_1$
is 
\begin{eqnarray}
\frac{T_2}{T_1}
=\frac{\kappa_z\kappa_{tz}}{\kappa_{zz}}\approx 
0.3\xi^2.
\label{T2/T1}
\end{eqnarray} 
with the dimensionless
quantity 
\begin{eqnarray}	\xi=\frac{\lambda}{L}.
\label{Q}
\end{eqnarray}
Here, $\lambda$ is
the mean free path 
of charged energetic
particles and $L$ is
the characteristic length 
of adiabatic focusing.
Obviously, the dimensionless
quantity $\xi$ determines
the relative 
importance of $T_2$ to 
$T_1$. 
From Equations (\ref{T2/T1})
and (\ref{Q}), we can find that 
if $\xi\gtrsim1$, 
adiabatic focusing effect
need to be taken into account. 
If  $\xi\ll 1$,
the adiabatic focusing effect
should be ignored. 

%\subsection{Relative
%importance of solar wind
%and adiabatic focusing  effects}
%\label{Exploring the relative
%$importance of adiabatic focusing to 
%solar wind effects}

In Subsection 
\ref{Exploring the 
importance of background plasma 
wind effect in the parallel 
diffusion}, 
the influence of  
solar wind 
on along-field diffusion 
is explored. 
In the condition with 
strong solar wind 
effect, 
$\alpha_1\sim \alpha_2\gg 1$,
considering the formulas of 
the solar wind effect $T_3$ and 
adiabatic focusing effect $T_2$, 
we have
\begin{eqnarray}
\frac{T_3}{T_2}=
\frac{27}{7}\frac{V^2L^2}
{u\omega\lambda^3}.
\end{eqnarray}
Obviously, the following dimensionless 
quantity can be found
\begin{eqnarray}
\gamma=\frac{V^2L^2}
{u\omega\lambda^3}
=\frac{V^2}
{u\omega\lambda}\xi^2. 
\end{eqnarray}
For typical parameter values
in Section \ref{The 
diffusion coefficients 
including solar wind effects},
thedimensionless quantity 
$\gamma\approx 0.3$. 
Thus,
if $\xi \sim 1$ is satisfied, 
both the solar wind effect and 
the adiabatic focusing effect
should be considered.

\section{The influence of
solar wind effect on	
cross-field  diffusion}
\label{The mean square
displacement definition of 
perpendicular diffusion  
along x direction}

Next, based on Equation 
(\ref{B}), we investigate 
the solar wind effect on 
eneregtic particle
cross-field diffusion.   

\subsection{The integrating form 
of the solar wind 
effect, $\eta$}
\label{The integrating form 
of stellar wind effect B}

From Equation (\ref{B}),
the solar wind effect
on the cross-field diffusion
can be written as
\begin{eqnarray}
\eta=\int_{-x_0}^{x_0}\dee xx
V_ xF_b	
-\int_{-x_0}^{x_0} \dee x xF_b	
\int_{-x_0}^{x_0} \dee x V_x F_b
\label{B int form}
\end{eqnarray}
From Figure \ref{fig1}, the following formula can be found
\begin{eqnarray}
V_\perp=V\sin\psi.
\end{eqnarray}
In this article, for the sake of
simplicity, we assume that
$V_\perp\sim
V_x$ and the solar wind
speed $V$ is constant.
Thus, Equation (\ref{B int form})
becomes
\begin{eqnarray}
\eta=V\int_{-x_0}^{x_0} \dee x x
F_b\sin\psi	
-V\int_{-x_0}^{x_0} \dee x x F_b	
\int_{-x_0}^{x_0} \dee x F_b\sin\psi.
\label{B int form-2}
\end{eqnarray}
Obviously,
the angle $\psi$ varies depending on
the point $A$ and 
obeys the following formula
on the ecliptic plane 
\citep[e.g.,][]{QiWu2018}
\begin{eqnarray} 
\sin\psi=
\frac{\omega r}{V}
\left(\frac{\omega^2 r^2}{V^2}+1\right)^{-\frac{1}{2}}.
\label{sin psi}
\end{eqnarray}
Inserting the latter formula
into Equation 
(\ref{B int form-2}), we have
\begin{eqnarray}
\eta=V\int_{-x_0}^{x_0} \dee xx
F_b\frac{\omega r}{V}
\left(\frac{\omega^2 r^2}{V^2}+1\right)^{-\frac{1}{2}}	
-V\int_{-x_0}^{x_0} \dee xx F_b	
\int_{-x_0}^{x_0} \dee x F_b
\frac{\omega r}{V}
\left(\frac{\omega^2 r^2}{V^2}+1\right)^{-\frac{1}{2}}.
\label{B int form-q}
\end{eqnarray}
To proceed, 
for simplicity, we assume
that the distribution
function $F_b$ is approximately
constant. Thus, 
the formula $\int_{-x_0}^{x_0} \dee xx F_b=0$ holds, and 
Equation (\ref{B int form-q})
becomes
\begin{eqnarray}
\eta=V\int_{-x_0}^{x_0} \dee xx
F_b\frac{\omega r}{V}
\left(\frac{\omega^2 r^2}{V^2}+1\right)^{-\frac{1}{2}}.
\label{B int form-q-2}	
\end{eqnarray}

From the equation of 
normal line, $y'=-x'/k+r_0$,
we can obtain the formulas
$\dee x=\sqrt{1+(\dee y'/\dee x')^2}
\dee x'=\sqrt{1+1/k^2}\dee x'$ and $x=\sqrt{1+(\dee 'y/\dee x')^2}
x'=\sqrt{1+1/k^2}x'$,
Equation 
(\ref{B int form-q-2}) becomes
\begin{eqnarray}
\eta=V
\left(1+\frac{1}{k^2}\right)
\int_{-0.1r_0}^{0.1r_0} \dee x' x'F_b
\frac{\omega r}{V}
\left(\frac{\omega^2 r^2}{V^2}+1\right)^{-\frac{1}{2}},
\label{B int form-q-3}
\end{eqnarray}
where we set
$x_0=0.1r_0/\cos\psi$.
Thus, 
from the normalization
condition 
$\int_{-0.1r_0}^{0.1r_0} \dee x'F_b=1$,
we can find
\begin{eqnarray}
	F_b=\frac{5}{r_0}. 
	\label{Fc=10/r0}
\end{eqnarray}
Accordingly, Equation
(\ref{B int form-q-3})
becomes
\begin{eqnarray}
\eta=V\frac{5}{r_0}
	\left(1+\frac{1}{k^2}\right)
	\int_{-0.1r_0}^{0.1r_0} \dee x' x'
	\frac{\omega r}{V}
	\left(\frac{\omega^2 r^2}{V^2}+1\right)^{-\frac{1}{2}}.
	\label{B int form-q-4}
\end{eqnarray}
Using the dimensionless 
quantity
$\alpha_1=V/(\omega r_0)$,
we can rewrite Equation 
(\ref{B int form-q-3}) as
\begin{eqnarray}
\eta=V\frac{5}{r_0}
\left(1+\frac{1}{k^2}\right)
\int_{-0.1r_0}^{0.1r_0} \dee x' x'
\left(\alpha_1^2
\frac{r_0^2}{r^2}+1\right)
^{-\frac{1}{2}}.	
\label{B int form-q-4}
\end{eqnarray}
The latter equation is 
very complex, so it is not 
easy to be evaluated. 
However, the relation
$r_0/r\sim 1$ 
approximately holds because
the integral interval
is small enough, 
so that 
Equation 
(\ref{B int form-q-4})
can be qualitatively explored
for $\alpha_1\gg1$ and 
$\alpha_1\ll1$.

\subsection{The condition $\alpha_1\gg1$}

For $\alpha_1\gg1$, 
with $r=\sqrt{x'^2+
\left(-x'/k+r_0\right)^2}$,
Equation (\ref{B int form-q-4}) 
is simplified as
\begin{eqnarray}
\eta=10\frac{\omega}{r_0}
\frac{1+k^2}{k^2}
\int_0^{0.1r_0} \dee x' x'
\sqrt{x'^2+\left(-x'/k+
	r_0\right)^2}.
\end{eqnarray}
After a lengthy mathematical
performance, the latter equation becomes 
\begin{eqnarray}
	&&\eta=\frac{10}{3}\omega r_0^2
	\frac{\left(1+k^2\right)^{3/2}}{k^3}
	\Bigg\{
	\left[\left(0.1-\frac{k}{1+k^2}\right)^2+\frac{k^4}{\left(1+k^2\right)^2}\right]^{3/2}
	-\left[\frac{k^2}{\left(1+k^2\right)^2}
	+\frac{k^4}{\left(1+k^2\right)^2}\right]^{3/2}
	\Bigg\}.
\end{eqnarray}
In the following, we 
evaluate the solar wind effect
on cross-field diffusion, $\eta$, for 
$\alpha_2=|k|\to \infty$ and 
$\alpha_2=|k|\to 0$, 
respectively.

\subsubsection{
The condition $\alpha_2=|k|\to \infty$}
\label{alpha2 to infty}

For the condition
\begin{eqnarray}
\alpha_2=|k|\to \infty,
\label{alpha2 to infty for cross}
\end{eqnarray}
we can easily
obtain
\begin{eqnarray}
	&&\eta\approx1.5\omega r_0^2.
\end{eqnarray}
Comparing the latter formula 
with $\kappa_{xx}=u\lambda_\perp/3$ we
obtain
\begin{eqnarray}
\frac{\eta}{\kappa_{xx}}
=4.5\frac{\omega r_0^2}{u\lambda_\perp}
\sim \frac{\omega r_0^2}{u\lambda_\perp}.
\end{eqnarray}
Obviously, the following
dimensionless quantity
determines the relative importance
of solar wind effect on
cross-field diffusion
\begin{eqnarray}
\delta_1=\frac{\omega 
r_0^2}{u\lambda_\perp}.
\end{eqnarray}
In the condition
(\ref{alpha2 to infty for cross}), 
the  point $A$ should be
close to the sun. Thus, 
the distance $r_0$ 
is much less than 1 AU,
i.e., $r_0\ll 10^{11}$ m. 
In addition, for the typical
parameter values
$\omega$, $u$, and 
$\lambda_\perp$ 
in Section
\ref{The diffusion coefficients including solar wind effects},
we have 
\begin{eqnarray}
\delta_1\ll1,
\end{eqnarray}
which
indicates the solar wind effect
on the along-field diffusion,
$\eta$,
may be ignored. 
 
\subsubsection{The condition
$\alpha_2\to 0$}

For $\alpha_2\to 0$, 
we can obtain
\begin{eqnarray}
&&\eta\approx
0.1\frac{\omega r_0^2}{k^3}.
\end{eqnarray}
Comparing the term $\eta$ and
the cross-field diffusion coefficient $\kappa_{xx}$
yields
\begin{eqnarray}
&&\frac{\eta}{\kappa_{xx}}
=0.3
\frac{\omega r_0^2}
{u\lambda_\perp k^3}.
\end{eqnarray}
It is obvious that the following 
dimensionless quantity determines
the relative importance of
solar wind to
cross-field diffusion 
\begin{eqnarray}
\delta_2=\frac{\omega r_0^2}{u\lambda_\perp k^3}.
\end{eqnarray}
For the typical parameter values
$u$, $\omega$, $\lambda_\perp$
in Section 
\ref{The diffusion coefficients including solar wind effects},
we can obtain 
\begin{eqnarray}
\delta_2\sim\left(
10^{-11}r_0
\right)^2\frac{1}{k^3}. 
\end{eqnarray}
For $\delta_2\ge 1$, 
the relation 
$k\le \left(10^{-11}r_0\right)
^{2/3}$
holds. However, the condition
$\alpha_1\gg1$ corresponds to
$r_0\ll V/\omega\sim 10^{11}$ m,
which denotes that 
the point $A$ is in the inner
heliosphere. 
The condition $\alpha_1\gg1$
is contradictory to $\alpha_2
\to 0$ which indicates the point 
$A$ in the outer heliosphere. 
Therefore, the solar wind 
effect on cross-field diffusion,
$\eta$, may be ignored.

\subsection{The condition
$\alpha_1\ll 1$}

If $\alpha_1\ll 1$, 
with $\int_{-0.1r_0}^{0.1r_0} 
\dee x F_b=1$,
Equation (\ref{B int form-q})
becomes
\begin{eqnarray}
\eta=&&V\int_{-0.1r_0}^{0.1r_0} 
\dee x  
xF_b
-V\int_{-0.1r_0}^{0.1r_0} \dee x 
xF_b=0.
\end{eqnarray}
Obviously, for this case,  
the solar wind effect, $\eta$,
might not be taken into 
account. 

%%%%%%%%%%%%%%%%%%%%%%%%%%%%%%%%%%%%%%%%%%%%%%%%%%%%%%%%%%%%%%%%%%%%%%%%%%%%%%%%%%%%%%%%%%%%%
\section{SUMMARY AND CONCLUSION}
\label{SUMMARY AND CONCLUSION}

In this paper, 
starting from the focusing equation, 
we derive the formulas of the
mean square displacement definitions
of along- and cross-field diffusion coefficients, 
$\kappa_{zz}^{DV}$
and $\kappa_{xx}^{DV}$,
respectively. 
It is demonstrated 
that 
$\kappa_{zz}^{DV}$ includes 
solar wind effect, i.e.,  
$\zeta=\left\langle zV_z\right\rangle
-\left\langle z\right\rangle
\left\langle V_z\right\rangle$,
and $\kappa_{xx}^{DV}$ 
contains 
$\eta=\left\langle xV_x \right\rangle
-\langle x \rangle\langle V_x \rangle$.

For different limits of 
dimensionless quantities
$\alpha_1=V/(\omega r_0)$ and $\alpha_2=|k|$,
the relative importance of 
the solar wind effect 
on along-field diffusion, $\zeta$,
is expored, and corresponding
dimensionless quantities
are obtained, which   
are summarized
in Table \ref{The cases for parallel diffusion with plasma wind effect}. 
For the condition 
$\alpha_1\sim \alpha_2\gg 1$,
we find that when the point $A$
is close to the sun, 
the relative 
importance of solar wind effect
on along-field diffusion, $\zeta$, 
should be taken into account. 
In this condition,
we find that  
when $\xi=\lambda/L\ge 1$,
the adiabatic focusing effects
need also to be considered. 
Next,   
the relative importance of
solar wind effect on
cross-field diffusion, $\eta$,
is investigated 
in several extreme
conditions, we find that  
the solar wind effect
on cross-field diffusion,
$\eta$, might be ignored. 

The results obtained in this paper
have certain significance 
in the transport of energetic
particles
not only in 
the heliosphere, but also 
in many other scenarios, 
such as, planetary 
magnetoshere and ionoshere, 
intersteller space, 
the spaces close to 
neutron stars, 
supernova remanents and so on.
It is possible that 
in some conditions,  
the solar wind
effect on energetic particle diffusion
in the heliosphere is not important, but 
the background plasma
speed effect on energetic
particle diffusion 
in other scenarios
is not ignorable.  

In this article, 
the results we get
are not very conclusive. 
We only perform the
exploration in
some extreme conditions.  
In addition,  
we set some
paramters with 
typical values as shown in 
Section 
\ref{The diffusion coefficients including solar wind effects},
which are standard 
in $1$ AU for energetic 
particles. However, 
using the typical parameter values,   
we discuss solar wind effect
on diffusion coefficients
in some special conditions,
e.g., $r_0\ll 1$ AU,
in which the typical parameter
values may be not appropriate
in the condition. 
In the future, we will 
investigate solar wind effect 
on diffusion coefficients 
in more general conditions,
using more reasonalble typical
parameter values in special
conditions.  

Other physical effects,
e.g., 
solar wind of spatial derivative, 
momentum transport, etc., 
on both spatial diffusion 
and drift coefficients,
may also be investigated. 
In past decades, the non-diffusion,
i.e., subdiffusion
and superdiffusion,
has gained more and more interest
due to its general applications 
in numerous research field.
We may also investigate  
the influence of 
the solar wind and
adiabatic focusing
on non-diffusion of energetic
particles.

\begin{acknowledgments}
The authors thank the anonymous referee for their valuable
comments.
We are partly supported by the Shenzhen Science and Technology 
Program under Grant No. JCYJ20210324132812029, and by
grants NNSFC 42074206 and NNSFC 42150105.
The work was supported by the National Key R\&D program of China 
No.2021YFA0718600 and No.2022YFA1604600, and by Shenzhen Key
Laboratory Launching Project (No. ZDSYS20210702140800001). 
The work was also supported by the Strategic Priority Research
Program of Chinese Academy of Sciences, Grant No. XDB 41000000.
\end{acknowledgments}

\renewcommand{\theequation}{\Alph{section}-\arabic{equation}}
\setcounter{equation}{0}  % reset counter
\begin{appendices}
\section{The focusing equation
satisfying particle number 
conservation law} 
\label{The focusing equation
	satisfying particle number 
	conservation law}

The focusing equation, which has been 
broadly used to research charged particle
transport, is as follow
\begin{eqnarray}
	\frac{\partial{f}}{\partial{t}}=&&\nabla\cdot\left(\kappa_{\perp}\cdot\nabla f\right)-
	\left(u\mu \hat{\bm{b}}+\bm{V}\right)\cdot
	\nabla f+\frac{\partial{}}{\partial{\mu}}
	\left(D_{\mu\mu}\frac{\partial{f}}{\partial{\mu}}\right)
	+\frac{dp}{dt}\frac{\partial{f}}{\partial{p}}+\frac{d\mu}{dt}\frac{\partial{f}}{\partial{\mu}}
	\label{old focusing equation}
\end{eqnarray}
with
\begin{eqnarray}
	&&\frac{dp}{dt}=p\Bigg[\frac{1-\mu^2}{2}\left(\nabla\cdot \bm{V}-\hat{\bm{b}}\hat{\bm{b}}
	:\nabla \bm{V}
	\right)
	+\mu^2\hat{\bm{b}}\hat{\bm{b}}:\nabla \bm{V}
	\Bigg]
	\label{dp/dt},\\
	&&\frac{d\mu}{dt}=\frac{1-\mu^2}{2}\Bigg[-\frac{u}{L}-
	\mu\left(\nabla\cdot \bm{V}-3\hat{\bm{b}}\hat{\bm{b}}
	:\nabla \bm{V}\right)\Bigg].
	\label{dp/dmu}	
\end{eqnarray}
Here, $L=\left(\hat{\bm{b}}\cdot\nabla\ln B\right)^{-1}$.
However, this formulation 
does not satisfy particle number
conservation law
which is one of the most important
physical laws. Therefore,
we have to find a focusing equation
which satisfies the particle number
conservation law. 

For spherical system $(p,\theta, \phi)$, the nabla operator of momentum
space is 
\begin{eqnarray}
\nabla_p f=\frac{\partial{f}}{\partial{p}}
\bm{e}_p
+\frac{1}{p}\frac{\partial{f}}{\partial{\theta}}\bm{e}_\theta
+\frac{1}{p\sin \theta}\frac{\partial{f}}{\partial{\phi}}\bm{e}_\phi.
	\label{nabla f for spherical sysem}
\end{eqnarray}
In this paper, we only consider gyrotropic case, and 
Equation (\ref{nabla f for spherical sysem}) becomes
\begin{eqnarray}
\nabla_p f=\frac{\partial{f}}{\partial{p}}\bm{e}_p
	+\frac{1}{p}\frac{\partial{f}}{\partial{\theta}}\bm{e}_\theta.
	\label{nabla gyrotropic f for spherical sysem}
\end{eqnarray}
Because $\mu=\cos \theta$, we have 
\begin{eqnarray}
	\frac{\partial{}}{\partial{\theta}}
	=\frac{\partial{}}{\partial{\mu}}\frac{\partial{\mu}}{\partial{\theta}}
	=-\sin \theta \frac{\partial{}}{\partial{\mu}}.
	\label{theta-mu}
\end{eqnarray}
Inserting the latter equation into Equation 
(\ref{nabla gyrotropic f for spherical sysem}) gives
\begin{eqnarray}
	\nabla_p f=\frac{\partial{f}}{\partial{p}}\bm{e}_p
	-\sin \theta
	\frac{1}{p}\frac{\partial{f}}{\partial{\mu}}
	\bm{e}_\theta.
	%\label{nabla gyrotropic f for spherical sysem for mu}
\end{eqnarray}
To proceed, we have 
\begin{eqnarray}
	\bm {a}\cdot \nabla_p f=\left(a_p\bm{e}_p+a_\theta\bm{e}_\theta \right)\cdot\left(\frac{\partial{f}}{\partial{p}}\bm{e}_p
	-\frac{\sin \theta}{p}\frac{\partial{f}}{\partial{\mu}}
	\bm{e}_\theta \right)
	=a_p\frac{\partial{f}}{\partial{p}}-a_\theta\frac{\sin \theta}{p}\frac{\partial{f}}{\partial{\mu}}
\end{eqnarray}
With the following formula 
\begin{eqnarray}
	\bm {a}\cdot \nabla_p f
	=\frac{dp}{dt}\frac{\partial{f}}{\partial{p}}
	+\frac{d\mu}{dt}\frac{\partial{f}}{\partial{\mu}},
\end{eqnarray}
we obtain 
\begin{eqnarray}
	&&\frac{dp}{dt}=a_p,\\
	&&\frac{d\mu}{dt}=-a_\theta\frac{\sin \theta}{p}.
\end{eqnarray}
That is, the latter formulas can be 
rewritten as 
\begin{eqnarray}
	&&a_p=\frac{dp}{dt},
	\label{ap}\\
	&&a_\theta=-\frac{d\mu}{dt}\frac{p}{\sin \theta}.
	\label{a theta}
\end{eqnarray}
Considering Equations (\ref{dp/dt}) and (\ref{dp/dmu}), we can find that 
\begin{eqnarray}
	&&a_p=p\Bigg[\frac{1-\mu^2}{2}\left(\nabla\cdot \bm{V}-\hat{\bm{b}}\hat{\bm{b}}
	:\nabla \bm{V}
	\right)
	+\mu^2\hat{\bm{b}}\hat{\bm{b}}:\nabla \bm{V}
	\Bigg]
	\label{a-p},\\
	&&a_\theta=\frac{p}{\sin \theta}
	\frac{1-\mu^2}{2}\Bigg[\frac{u}{L}+
	\mu\left(\nabla\cdot \bm{V}-3\hat{\bm{b}}\hat{\bm{b}}
	:\nabla \bm{V}\right)\Bigg]
	\label{a-theta}. 
\end{eqnarray}
Inserting formulas
(\ref{ap}) and (\ref{a theta})
into the following equation
\begin{eqnarray}	\nabla_p\cdot\left(\bm{a}f\right)
=&&\frac{1}{p^2}\frac{\partial{}}{\partial{p}}
\left(p^2a_pf\right)
-\frac{1}{p}\frac{\partial{}}{\partial{\mu}}
\left(\sin\theta a_\theta f\right),
\end{eqnarray}
we can obtain
\begin{eqnarray}
	\nabla_p \cdot \left(\bm {a}f\right)
	=&&\frac{1}{p^2}\frac{\partial{}}{\partial{p}}
	\Bigg\{p^3\Bigg[\frac{1-\mu^2}{2}\left(\nabla\cdot \bm{V}-\hat{\bm{b}}\hat{\bm{b}}
	:\nabla \bm{V}
	\right)
	+\mu^2\hat{\bm{b}}\hat{\bm{b}}:\nabla \bm{V}
	\Bigg]f\Bigg\}
	\nonumber\\
	&&
	+\frac{\partial{}}{\partial{\mu}}
	\Bigg\{
	\frac{1-\mu^2}{2}\Bigg[-\frac{u}{L}-
	\mu\left(\nabla\cdot \bm{V}-3\hat{\bm{b}}\hat{\bm{b}}
	:\nabla \bm{V}\right)\Bigg]f
	\Bigg\}.
\end{eqnarray}

The Fokker-Planck equation
is shown as
\begin{eqnarray}
	\frac{\partial{f}}{\partial{t}}
	+\nabla\cdot \left(\bm {v}f\right)
	+\nabla_p \cdot \left(\bm {a}f\right)
	=\nabla\cdot\left(\kappa\cdot \nabla f\right)
	+\nabla_p\cdot\left(\kappa\cdot \nabla_p f\right),
	\label{Fokker-Planck equation in Appendix}
\end{eqnarray}
which satisfies the particle number
conservation law. 
By considering the latter formula
and the following fomulas
\begin{eqnarray}
&&\bm {v}=u\mu \hat{\bm{b}}+\bm{V},\\     
	&&\nabla\cdot
	\left(\kappa\cdot \nabla f\right)
	=\nabla\cdot
	\left(\kappa_\perp\cdot \nabla f\right),\\            
	&&\nabla_p\cdot\left(\kappa\cdot \nabla_p f\right)
	=\frac{\partial{}}{\partial{\mu}}
	\left(D_{\mu\mu}\frac{\partial{f}}{\partial{\mu}}\right),
\end{eqnarray}
Equation (\ref{Fokker-Planck equation in Appendix}) becomes 
\begin{eqnarray}
	&&\frac{\partial{f}}{\partial{t}}
	=\nabla\cdot\left(
	\kappa_\perp\cdot \nabla f\right)
	-\nabla\cdot \left[\left(u\mu \hat{\bm{b}}+\bm{V}\right)f\right]
	+\frac{\partial{}}{\partial{\mu}}
	\left(D_{\mu\mu}\frac{\partial{f}}{\partial{\mu}}\right)
	\nonumber\\
	&&
	+\frac{1}{p^2}\frac{\partial{}}{\partial{p}}
	\Bigg\{p^3\Bigg[\frac{1-\mu^2}{2}\left(\nabla\cdot \bm{V}-\hat{\bm{b}}\hat{\bm{b}}
	:\nabla \bm{V}
	\right)
	+\mu^2\hat{\bm{b}}\hat{\bm{b}}:\nabla \bm{V}
	\Bigg]f\Bigg\}
	\nonumber\\
	&&
	+\frac{\partial{}}{\partial{\mu}}
	\Bigg\{
	\frac{1-\mu^2}{2}
	\Bigg[-\frac{u}{L}-
	\mu\left(\nabla\cdot \bm{V}-3\hat{\bm{b}}\hat{\bm{b}}
	:\nabla \bm{V}\right)
	\Bigg]f
	\Bigg\},
	%\label{equation satisfying particle number conservation}
\end{eqnarray}
which is our starting point of this 
paper.

In fact, if the incompressible
condition   
\begin{eqnarray}
	\nabla\cdot\bm{v}+\nabla_p\cdot\bm{a}=0
	%\label{impressible condition}
\end{eqnarray}
holds, Equation (\ref{old focusing equation}) can also 
satisfy particle number
conservation. Now, the specific 
form of the incompressible
condition is derived. 
With the divergence formula 
\begin{eqnarray}
	\nabla\cdot\bm{A}=\frac{1}{r^2}\frac{\partial{}}{\partial{r}}
	\left(r^2f_r\right)
	+\frac{1}{r\sin \theta}\frac{\partial{}}{\partial{\theta}}
	\left(\sin\theta f_\theta\right)
	+\frac{1}{r\sin \theta}
	\frac{\partial{f_\phi}}{\partial{\phi}},
\end{eqnarray}
for gyrotropic case the latter formula
becomes
\begin{eqnarray}
	\nabla\cdot\bm{A}=\frac{1}{r^2}\frac{\partial{}}{\partial{r}}
	\left(r^2f_r\right)
	+\frac{1}{r\sin \theta}\frac{\partial{}}{\partial{\theta}}
	\left(\sin\theta f_\theta\right)
\end{eqnarray}
Using $\mu=\cos\theta$, the latter 
formula can be rewritten as
\begin{eqnarray}
	\nabla\cdot\bm{A}
	=\frac{1}{r^2}\frac{\partial{}}{\partial{r}}
	\left(r^2f_r\right)
	-\frac{1}{r}\frac{\partial{}}{\partial{\mu}}
	\left(\sin\theta f_\theta\right).
\end{eqnarray}
In addition, the divergence formula
for velocity is shown as
\begin{eqnarray}
	\nabla\cdot\bm{v}=\nabla\cdot\left(u\mu \hat{\bm{b}}+\bm{V}\right)
	=\nabla\cdot\bm{V}.
	\label{divergence formula
		for velocity}
\end{eqnarray}
Similarly, 
with Equations (\ref{ap})
and (\ref{a theta}),
the divergence formula 
for acceleration can be obtained
\begin{eqnarray}
	\nabla_p\cdot\bm{a}=&&\frac{1}{p^2}\frac{\partial{}}{\partial{p}}
	\left(p^2a_p\right)
	-\frac{1}{p}\frac{\partial{}}{\partial{\mu}}
	\left(\sin\theta a_\theta\right)
	\nonumber\\
	&&
	=\frac{1}{p^2}\frac{\partial{}}{\partial{p}}
	\Bigg\{p^3\Bigg[\frac{1-\mu^2}{2}\left(\nabla\cdot \bm{V}-\hat{\bm{b}}\hat{\bm{b}}
	:\nabla \bm{V}
	\right)
	+\mu^2\hat{\bm{b}}\hat{\bm{b}}:\nabla \bm{V}
	\Bigg]\Bigg\}
	\nonumber\\
	&&
	-\frac{1}{p}\frac{\partial{}}{\partial{\mu}}
	\Bigg\{\sin\theta \frac{p}{\sin \theta}
	\frac{1-\mu^2}{2}\Bigg[\frac{u}{L}+
	\mu\left(\nabla\cdot \bm{V}-3\hat{\bm{b}}\hat{\bm{b}}
	:\nabla \bm{V}\right)\Bigg]\Bigg\}
\end{eqnarray}
To continue, the latter formula
can be simplified as
\begin{eqnarray}
\nabla_p\cdot\bm{a}
=\nabla\cdot\bm{V}
+\left(3\mu^2-1\right)\hat{\bm{b}}\hat{\bm{b}}:\nabla \bm{V}
+\frac{u\mu}{L}.
\label{nabla a}
\end{eqnarray}
Combining Equations 
(\ref{divergence formula
	for velocity}) and 
(\ref{nabla a}) yields 
\begin{eqnarray}
	\nabla\cdot\bm{v}+\nabla_p\cdot\bm{a}
	=2\nabla\cdot\bm{V}
	+\left(3\mu^2-1\right)\hat{\bm{b}}\hat{\bm{b}}:\nabla \bm{V}
	+\frac{u\mu}{L}=0
	%\label{impressible condition}
\end{eqnarray}
Thus, the condition of particle number
conservation can be rewritten as
\begin{eqnarray}
-\frac{u\mu}{L}
=2\nabla\cdot\bm{V}
	+\left(3\mu^2-1\right)\hat{\bm{b}}\hat{\bm{b}}:\nabla \bm{V}
	%\label{impressible condition}
\end{eqnarray}
If the latter condition 
is satisfied, Equation 
(\ref{old focusing equation})
also satisfies the particle number
conservation law.

%%%%%%%%%%%%%%%%%%%%%%%%%%%%%
%%%%%%%%%%%%%%%%%%%%%%%%%%%%%
%\renewcommand{\theequation}{\Alph{section}-\arabic{equation}}
\setcounter{equation}{0}  % reset counter
%\begin{appendices}
%\appendix

\section{The focusing equation  without tensor operation}
\label{The focusing equation  without tensor operation}

For mathematical tractability,  
the focusing equation can be 
rewritten as
\begin{eqnarray}
	&&\frac{\partial{f}}{\partial{t}}
	=\nabla\cdot
	\left(\kappa_\perp\cdot \nabla f\right)
	-\nabla\cdot \left[\left(u\mu \hat{\bm{b}}+\bm{V}\right)f\right]
	+\frac{\partial{}}{\partial{\mu}}
	\left(D_{\mu\mu}\frac{\partial{f}}{\partial{\mu}}\right)
	\nonumber\\
	&&
	+\frac{1}{p^2}\frac{\partial{}}{\partial{p}}
	\Bigg\{p^3\Bigg[\frac{1-\mu^2}{2}
	\left(\frac{\partial{V_{x}}}
	{\partial{x}}
	+\frac{\partial{V_{y}}}
	{\partial{y}}\right)
	+\mu^2\frac{\partial{V_z}}
	{\partial{z}}
	\Bigg]f\Bigg\}
	\nonumber\\
	&&
	+\frac{\partial{}}{\partial{\mu}}
	\Bigg\{
	\frac{1-\mu^2}{2}\Bigg[-\frac{u}{L}-
	\mu\left(\frac{\partial
		{V_{x}}}
	{\partial{x}}
	+\frac{\partial{V_{y}}}
	{\partial{y}}-2
	\frac{\partial{V_{z}}}
	{\partial{z}}\right)\Bigg]f
	\Bigg\},
	\label{equation satisfying particle 
		number conservation
		without tensor operation in Appendix}
\end{eqnarray}
with the following formulas
\begin{eqnarray}
	&&\bm{V}^{SW}=V_{z}^{SW}\hat{\bm{b}}+
	V_{x}^{SW}\hat{\bm{n}_x}+
	V_{y}^{SW}\hat{\bm{n}_y},\\
	&&\nabla=\frac{\partial}{\partial{z}}
	\hat{\bm{b}}+\frac{\partial}{\partial{x}}
	\hat{\bm{n}_x}+\frac{\partial}{\partial{y}}
	\hat{\bm{n}_y},\\
	&&\hat{\bm{b}}\cdot\nabla=\hat{\bm{b}}\cdot
	\left(\frac{\partial}{\partial{z}}
	\hat{\bm{b}}+\frac{\partial}{\partial{x}}
	\hat{\bm{n}_x}+\frac{\partial}{\partial{y}}
	\hat{\bm{n}_y}\right)
	=\frac{\partial}{\partial{z}},\\
	&&\nabla \cdot\hat{\bm{b}}=
	\left(\frac{\partial}{\partial{z}}
	\hat{\bm{b}}+\frac{\partial}{\partial{x}}
	\hat{\bm{n}_x}+\frac{\partial}{\partial{y}}
	\hat{\bm{n}_y}\right)\cdot\hat{\bm{b}}
	=\frac{\partial}{\partial{z}},\\
	&&\hat{\bm{b}}\cdot\bm{V}^{SW}
	=\hat{\bm{b}}\cdot\left(V_{z}^{SW}\hat{\bm{b}}+
	V_{x}^{SW}\hat{\bm{n}_x}+
	V_{y}^{SW}\hat{\bm{n}_y}\right)
	=V_{z}^{SW},\\
	&&\hat{\bm{b}}\hat{\bm{b}}:\nabla \bm{V}^{SW}=
	\left(\hat{\bm{b}}\cdot\nabla\right)
	\left(\hat{\bm{b}}\cdot\bm{V}^{SW}\right)
	=\frac{\partial{V_z^{SW}}}{\partial{z}},\\
	&&\nabla\cdot\bm{V^{SW}}=\left(\frac{\partial}{\partial{z}}
	\hat{\bm{b}}+\frac{\partial}{\partial{x}}
	\hat{\bm{n}_x}+\frac{\partial}{\partial{y}}
	\hat{\bm{n}_y}\right)
	\left(V_{z}^{SW}\hat{\bm{b}}+
	V_{x}^{SW}\hat{\bm{n}_x}+
	V_{y}^{SW}\hat{\bm{n}_y}\right)
	=\frac{\partial{V_{z}^{SW}}}{\partial{z}}
	+\frac{\partial{V_{x}^{SW}}}{\partial{x}}
	+\frac{\partial{V_{y}^{SW}}}{\partial{y}},\\
	&&\nabla\cdot \bm{V^{SW}}-\hat{\bm{b}}\hat{\bm{b}}
	:\nabla \bm{V}^{SW}
	=\frac{\partial{V_{x}^{SW}}}
	{\partial{x}}
	+\frac{\partial{V_{y}^{SW}}}
	{\partial{y}},\\
	&&\nabla\cdot\left(\bm{V}^{SW}F\right)
	=\left(\frac{\partial}{\partial{z}}
	\hat{\bm{b}}+\frac{\partial}{\partial{x}}
	\hat{\bm{n}_x}+\frac{\partial}{\partial{y}}
	\hat{\bm{n}_y}\right)
	\left(V_{z}^{SW}F\hat{\bm{b}}+
	V_{x}^{SW}F\hat{\bm{n}_x}+
	V_{y}^{SW}F\hat{\bm{n}_y}\right)
	\nonumber\\
	&&
	=\frac{\partial{\left(V_{z}^{SW}F\right)}}{\partial{z}}+\frac{\partial{\left(V_{x}^{SW}F\right)}}{\partial{x}}
	+\frac{\partial{\left(V_{y}^{SW}F\right)}}{\partial{y}},\\
	&&\nabla\cdot \bm{V^{SW}}-3\hat{\bm{b}}\hat{\bm{b}}
	:\nabla \bm{V}^{SW}
	=\frac{\partial{V_{x}^{SW}}}
	{\partial{x}}
	+\frac{\partial{V_{y}^{SW}}}
	{\partial{y}}-2
	\frac{\partial{V_{z}^{SW}}}
	{\partial{z}}.
\end{eqnarray}

\end{appendices}

%%%%%%%%%%%%%%%%%%%%%%%%%%%%%%%%%%%%%%%%%%%%%%%%%%%%%%
\clearpage

%%%%%%%%%%%%%%%%%%%%%%%%%%%%%%%%%%%%%%%%%%%%%%%%%%%%%%%%

\clearpage
\begin{table}[ht]
	\begin{center}
		\caption{The dimensionless 
			quantities determining
			the relative importance of solar wind effect
		on along-field diffusion}
		\label{The cases for parallel diffusion with plasma wind effect}
		\begin{tabular}{|c|c|c|}
			\hline 
			\multirow{1}{*}{Range}   & 
			\multicolumn{2}{c|}{
			\multirow{1}{*}
			{solar wind effect}} 
			%{Solar wind and} 
		    %\\
		    %& \multicolumn{2}{c|}{} &  
		    %adiabatic focusing effects 
		    \\
			\hline
			%%%%%%%%%%%%
			\multirow{1}{*}{$\alpha_1\gg\alpha_2\gg1$}   & \multirow{1}{*}
			{$A=|k|Vr_0$} & $\beta_1=\frac{|k|Vr_0}{u\lambda}$ 
			%& $\gamma_1=|k|\frac{Vr_0L^2}
			%{u\lambda^3}$
			\\
						\hline
%%%%%%%%%%%%
\multirow{2}{*}{$\alpha_2\gg\alpha_1\gg1$}   
& \multirow{2}{*}
{$A=-10\frac{V^2}
{\omega}\ln \frac{\omega r_0 |k|}{V}$} 
& $\beta_2=\frac{V^2}
{u\omega\lambda}$,  
%\multirow{2}{*}
%& $\gamma_2=\frac{V^2L^2}  
%{u\omega\lambda^3}$,  
\\
&    & $\beta_3=\frac{\omega r_0 |k|}{V}$
%& $\beta_3=\frac{\omega r_0 |k|}{V}$
\\
\hline
%%%%%%%%%%%%
\multirow{1}{*}{$\alpha_1\sim \alpha_2\gg1$}
& \multirow{1}{*}{$A=0.4\frac{V^2}{\omega}$} 
& $\beta_2=\frac{V^2}{u\omega\lambda}$ 
%& $\gamma_2=\frac{V^2L^2}{u\omega\lambda^3}$
\\
\hline

%%%%%%%%%%%%
\multirow{1}{*}{$\alpha_1\gg1$, $\alpha_2\ll1$}   
& \multirow{1}{*}{$A=|k|Vr_0$} 
& $\beta_1=\frac{|k|Vr_0}{u\lambda}$ 
%& $\gamma_1=|k|\frac{Vr_0L^2}{u\lambda^3}$
\\
\hline

%%%%%%%%%%%%
\multirow{1}{*}{$\alpha_2\ll1$, $\alpha_1 \gg 1$}   
& \multirow{1}{*}{$A=10\frac{V^2}
{\omega}$} 
& $\beta_2=\frac{V^2}
{u\omega\lambda}$ 
%& $\gamma_2=\frac{V^2L^2}{u\omega\lambda^3}$
\\
\hline

%%%%%%%%%%%%
\multirow{1}{*}{$\alpha_1\ll1$, $\alpha_2\ll1$}   
& \multirow{1}{*}{$A=|k|\frac{V^2}{\omega}$} 
& $\beta_4=\frac{|k|V^2}
{u\omega\lambda}$ 
%& $\gamma_4=|k|\frac{V^2L^2}{u\omega\lambda^3}$ 
\\
\hline
		\end{tabular}
	\end{center}
\end{table}

\end{document}